\documentclass[aps,amsfonts,pre,twocolumn,superscriptaddress,showpacs,eqsecnum]{revtex4-1}
\usepackage{epsfig,amsmath,amssymb,bm,epsf,graphics,psfrag,verbatim,subfigure}
\def\s{\mu}
\def\b{\kappa}
\def\sm{\mu_m}
\def\bm{\kappa_m}
\def\lm{\lambda_m}
\def\ss{\mu_s}
\def\bs{\kappa_s}
\def\vr{\mathbf{r}}
\def\vR{\mathbf{R}}
\def\vu{\mathbf{u}}
\def\uvr{\hat{\mathbf{r}}}

\def\uvepara{\mathbf{e}}
\def\uveperp{\mathbf{e}^{\perp}}
\def\tPPV{\tilde{V}}

\def\tGF{\tilde{G}}
\def\tGM{\tGF^{(m)}}
\def\tTM{\tilde{T}}
\def\tT{\tilde{T}}
\def\PCF{\Prob_{\textrm{CF}}}
\def\tIM{\tilde{I}}
\def\GM{\GF^{(m)}}

\def\Prob{p}
\def\Pb{p_{\textrm{b}}}
\def\PCF{p_{\textrm{CF}}}

\def\DM{\mathbf{D}^{(m)}}
\def\vu{\mathbf{u}}
\def\vq{\mathbf{q}}

\def\vB{\mathbf{B}}

\def\PPV{\mathbf{V}}
\def\GF{\mathbf{G}}
\def\TM{\mathbf{T}}

\def\btwo{\b_c}

\def\bmr{b_m}
\def\lmr{l_m}
\def\br{b}

\def\bmr{b_m}
\def\lmr{l_m}
\def\br{b}
\def\deltap{\delta p}
\def\dsm{\delta \sm}
\def\dbm{\delta \bmr}
\def\dlm{\delta \lmr}
\def\lb{l_{\textrm{bend}}}
\def\a{a}

\def\DMT{\mathbf{D}}
\def\vv{\mathbf{v}}
\def\MM{\mathbf{M}}
\def\IM{\mathbf{I}}

\def\ve{\mathbf{e}}
\def\kbar{\overline{\kappa}}

\def\lone{\ell_1}
\def\ltwo{\ell_2}
\def\lthree{\ell_3}
\def\lfour{\ell_4}

\begin{document}
\title{Effective Medium Theory of Filamentous Triangular Lattice}
\author{Xiaoming Mao}
\affiliation{Department of Physics and Astronomy, University of
Pennsylvania, Philadelphia, PA 19104, USA }
\affiliation{Department of Physics, University of
Michigan, Ann Arbor, MI 48109-1040, USA }
\author{Olaf Stenull}
\affiliation{Department of Physics and Astronomy, University of
Pennsylvania, Philadelphia, PA 19104, USA }
\author{T. C. Lubensky}
\affiliation{Department of Physics and Astronomy, University of
Pennsylvania, Philadelphia, PA 19104, USA }

\date{\today}

\begin{abstract}
We present an effective medium theory that includes bending as
well as stretching forces, and we use it to calculate
mechanical response of a diluted filamentous triangular
lattice. In this lattice, bonds are central-force springs, and
there are bending forces between neighboring bonds on the same
filament.  We investigate the diluted lattice in which each
bond is present with a probability $\Prob$. We find a rigidity
threshold $\Pb$ which has the same value for all positive
bending rigidity and a crossover characterizing bending-,
stretching-, and bend-stretch coupled elastic regimes
controlled by the central-force rigidity percolation point
at $\PCF \simeq 2/3$ of the lattice when fiber bending rigidity
vanishes.

\end{abstract}


\maketitle

\section{Introduction}

Random elastic networks provide attractive and realistic models
for the mechanical properties of materials as diverse as
randomly packed spheres \cite{Liu1998,Wyart2005a,LiuNag2010a},
network glasses
\cite{Phillips1981,Thorpe1983,Phillips1985,HeTho1985,Thorpe2000},
and biopolymer gels
\cite{Elson1988,Kasza2007,Alberts2008,JanmeySto1990,MacKintosh1995,Head2003,Wilhelm2003,Head2005,Storm2005,OnckVan2005,HuismanVan2007,Huisman2010}.
In their simplest form, these networks consist of nodes
connected by central-force (CF) springs to on average of $z$
neighbors. They become more rigid as $z$ increases, and they
typically exhibit a CF rigidity percolation transition
\cite{Feng1984,FengLob1984,Jacobs1995} from floppy clusters to
a sample spanning-cluster endowed with nonvanishing shear and
bulk moduli at a threshold $z=z_{\textrm{CF}}$ very close to
the Maxwell isostatic limit \cite{Maxwell1864,Calladine1978} of
$2d$, where $d$ is the spatial dimension, at which the number
of constraints imposed by the springs equals the number of
degrees of freedom of individual nodes. Generalized versions of
these networks, appropriate for the description of network
glasses \cite{Phillips1981,Thorpe1983} and biopolymer gels
\cite{MacKintosh1995,Head2003,Wilhelm2003}, include bending
forces favoring a particular angle between bonds (springs)
incident on a given node. For a given value of $z$, networks
with bending forces are more rigid than their CF-only
counterparts, and they exhibit a rigidity transition at
$z=z_{\textrm{b}} <z_{\textrm{CF}}$.

\begin{figure}%[t]
    \centering
     \includegraphics[width=.45\textwidth]{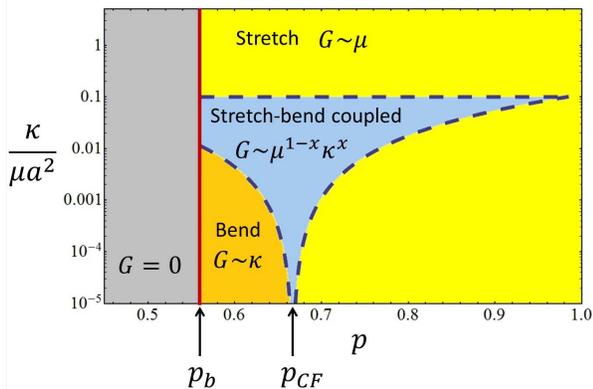}
        \caption{(Color online) Phase diagram of the diluted filamentous
        triangular lattice showing the central-force and bending rigidity thresholds,
        respectively, at $p = \PCF$ and $p=\Pb$, the bending-dominated regime at small $\kappa$
        in the vicinity of $\Pb$, the crossover bend-stretch regime near $\PCF$, and the
        stretching dominated regime at large $\kappa$. (Adapted from Ref.~\cite{BroederszMac2011})
        }
\label{FIG:PD}
\end{figure}

Though numerical calculations, including the pebble game
\cite{Jacobs1995,JacobsTho1996}, have provided much of our
knowledge about the properties of random elastic networks,
effective medium theories (EMTs)
\cite{Soven1969,Elliott1974,Feng1985,Garboczi1985,SchwartzSen1985}
have provided complementary analytical descriptions of CF
networks that are simple and at minimum qualitatively correct.
EMTs
\cite{Das2007,Heussinger2006,HeussingerFre2007a,BroederszMac2011,Das2012}
and heuristic approaches \cite{WyartMah2008} that describe both
bending and stretching forces have only recently been
developed. Here we present details of the derivation of a
bend-stretch EMT introduced in Ref.~\cite{BroederszMac2011} and
its application to a bond diluted triangular lattice, whose
maximum coordination number is $z_{\rm max}=6$. This lattice
has bending and stretching forces modeled on those of
biopolymer networks of filamentous semi-flexible polymers,
characterized by  one-dimensional stretching and bending moduli
$\mu$ and $\kappa$ \cite{MacKintosh1995,Head2003,Wilhelm2003},
respectively. Our EMT calculates the effective medium moduli
$\mu_m$ and $\kappa_m$ as a function of $\mu$ and $\kappa$ and
the probability $p=z/6$ that a bond is occupied. Both the EMT
bulk and shear moduli are proportional to $\mu_m$. When $\kappa
= 0$, our EMT reduces to that considered by others
\cite{Feng1985,Garboczi1985} and successfully predicts a
second-order CF rigidity threshold at $z_{\textrm{CF}}\simeq 4<
z_{\rm max}$ ($\PCF =2/3$ in EMT and of order $0.64$ to
$0.65$ under various numerical estimates
\cite{Sahimi1998,WangAdl1992,BroederszMac2011} ) with $\mu_m$
increasing linearly in $p-\PCF$ near $\PCF$ and approaching the
undiluted triangular-lattice value of $\mu$ at $p=1$. When
bending forces are introduced, our EMT predicts a second-order
rigidity threshold $p_b < p_{CF}$ for \emph{all} $\kappa>0$.
This qualitatively agrees with the results of an
alternative EMT in Ref.~\cite{Das2012}, although our theory
predicts $p_b\simeq 0.56$ in poorer agreement with the value
$p_b \simeq 0.44$ obtained in simulations than the value $p_b
\simeq 0.457$ predicted there. Near $p_b$ we find that $\mu_m
\sim \kappa (p-\Pb)$ for $\kappa/(\mu \a^2) \ll c_1 \approx
0.1$ and $\mu_m \sim \mu (p-\Pb)$ for $\kappa/(\mu \a^2)\gg
c_1$, where $\a$ is the lattice spacing. Near $\PCF$,
$\kappa$ is a relevant variable moving the system away from the
CF rigidity critical point to a broad crossover regime
\cite{WyartMah2008,BroederszMac2011} in which $\mu_m \sim
\kappa^{1/2} \mu^{1/2}$ as shown in the phase diagram of
Fig.~\ref{FIG:PD}. This crossover is analogous to that for the
macroscopic conductivity in a resistor network in which bonds
are occupied with resistors with conductance $\sigma_{>}$  with
probability $p$ and with conductance $\sigma_< < \sigma_>$ with
probability $1-p$ \cite{Straley1976}.

Though the model we study has both stretching and bending
forces, it differs in important ways from previously studied
models for network glasses
\cite{Phillips1981,Thorpe1983,Phillips1985,HeTho1985,Thorpe2000}
and for filamentous gels
\cite{MacKintosh1995,Head2003,Wilhelm2003,Head2005,Storm2005,OnckVan2005,HuismanVan2007,Huisman2010}.
The maximum coordination number for both of these systems is
less than or equal $2d$, and thus neither has a CF rigidity
transition for $p<1$ when there are no bending forces. As a
result neither exhibits the bend-stretch crossover region near
$\PCF$ that our model exhibits. Network glasses are well
modeled by a randomly diluted four-fold coordinated diamond
lattice in which there is a bending-energy cost, characterized
by a bending modulus $\kappa$, if the angle between any pair of
bonds incident on a site deviates from the tetrahedral angle of
$109.5 \deg$. The architecture of the undiluted diamond lattice
(with $z_{\rm max} =4 <2d = 6$) is such that its shear modulus
vanishes linearly with $\kappa$ \cite{HeTho1985} and elastic
response is nonaffine. When diluted, it exhibits a second-order
rigidity transition from a state with bending-dominated
nonaffine shear response to a state with no rigidity. As
dilution decreases, rigidity is still controlled by $\kappa$,
but response becomes less nonaffine.

Filamentous networks in two-dimensions are often described by
the Mikado model \cite{Head2003,Wilhelm2003,Head2005} in which
semi-flexible filaments of a given length $L$ are deposited
with random center-of-mass position and random orientation on a
two-dimensional plane and in which the points where two
filaments cross are joined in frictionless crosslinks. As in
our model, there is no energy cost for the relative rotation of
two rods about a crosslink, but there is an energy cost for
bending the rods at crosslinks. This model is characterized by
the ratio $\eta \equiv L/l_c$ of the filament length $L$ to the
average mesh size, i.e., the average crosslink separation $l_c
> a$ along a filament, where $a$ is the shortest distance between crosslinks. In
the limit $\eta\to \infty$, all filaments traverse the sample,
and the system has finite, $\kappa$-independent shear and bulk
moduli: There is effectively a CF rigidity transition at $z=4$
when $\eta$ is decreased from infinity. There is a transition
at $\eta=\eta_c \approx 5.9$ from a floppy to a rigid state
with nonaffine response~\cite{Head2003,LatvaTim2001b}, and
there is a wide crossover region between $\eta = \eta_c$ and
$\eta = \infty$ in which the shear modulus changes from being
bend dominated, nonaffine, and nearly independent of $\mu$ at
small $\eta$ to being stretch dominated, nearly affine, and
nearly independent of $\kappa$ at large $\eta$. Our EMT applied
to the kagome lattice \cite{MaoLub2011b}, whose maximum
coordination number like that of the Mikado model is four,
captures these crossovers. Interestingly, $3d$ lattices
composed of straight filaments with $z_{\rm max}=4$ exhibit
similar behavior~\cite{Stenull2011}. When filaments are bent,
however, elastic response in one case at least
\cite{Huisman2010} is more like that of the diluted diamond
lattice with the shear modulus vanishing with $\kappa$ even at
at large $L/l_c$ or $z$ near $4$.

External tensile stress (i.e., negative pressure) can cause a
floppy lattice to become rigid \cite{Alexander1998}.  Random
internal stresses can do so as well in a phenomenon called
tensegrity \cite{Calladine1978}. Thus a lattice with internal
stresses may have a lower rigidity threshold than the same
lattice with out internal stresses \cite{Huisman2010}. Systems
such as network glasses can exhibit two rigidity transitions
\cite{Thorpe2000,BoolchandTho2005}: a second-order transition
from a floppy to a rigid but unstressed state followed closely
by a first-order transition to a rigid but stressed state.
These effects are beyond the scope of EMT and will not be
treated.

The outline of our paper is as follows. Section~\ref{SEC:MODEL}
reviews properties of semi-flexible polymers and defines our
model for the harmonic elasticity of crosslinked semi-flexible
polymers on a triangular lattice; Sec.~\ref{SEC:CPA} sets up
our effective medium theory; Sec.~\ref{SEC:Results} presents
the results of this theory; and Sec.~\ref{SEC:DISCUSSION}
compares our EMT with other versions of bend-stretch EMTs and
summarizes our results.  There are four appendices:
App.~\ref{APP:COMP} derives the energy, which is critical to
our version of EMT, of a composite bent rod, App.~\ref{APP:DM}
presents the detailed form of the dynamical matrix,
and
App.~\ref{APP:Comp} provides a detailed comparison of our EMT
and that of Refs.~\cite{Das2007,Das2012}.

\section{Filamentous Polymers on a Triangular Lattice}\label{SEC:MODEL}
\subsection{Elastic Rods: Continuum and Discretized Energies}

Following previous work~\cite{Head2003,Wilhelm2003}, we model
individual filaments as homogeneous elastic rods characterized
by a stretching (or Young's) modulus $\mu$ and a bending
modulus $\kappa$. We restrict out attention to two dimensions.
The filament energy is thus,
\begin{eqnarray}\label{EQ:Rod}
    E = \frac{1}{2}\int_{0}^{L} ds \left[
        \mu \left(\frac{du(s)}{ds}\right)^2 + \kappa \left(\frac{d\theta(s)}{ds}\right)^2
    \right],
\end{eqnarray}
where $s$ is the arclength coordinate, $L$ is the unstretched
contour length of the polymer, and $u(s)$ and $\theta(s)$ are,
respectively, the longitudinal displacement and angle of the
unit tangent to the polymer at $s$. We treat this as a purely
mechanical model in which $\mu$ and $\kappa$ are fixed, and we
do not consider the entropic contributions to the energy that
arise from thermally induced transverse fluctuations of the
filaments \cite{MacKintosh1995,MarkoSig1995,Storm2005}. Three
length scales can be identified in this elastic energy. The
first is the contour length of the polymers, $L$. The second,
$\lb\equiv\sqrt{\kappa/\mu}$, characterizes the relative
strength of stretching and bending. For an elastic rod made of
a homogeneous material, $\lb$ is simply proportional to the
radius of the rod.  A third length, the mesh size $l_c$
characterizing the connectivity of the network, can be
identified for crosslinked polymer networks. The ratio $L/l_c$
is a measure of the connectivity of the lattice. Finite
filaments of length $L$ with this energy act like springs with
stretching spring constant $k_{||} = \mu/L$ and bending
constant $\kappa/L^3$.

\begin{figure}
\centerline{\includegraphics{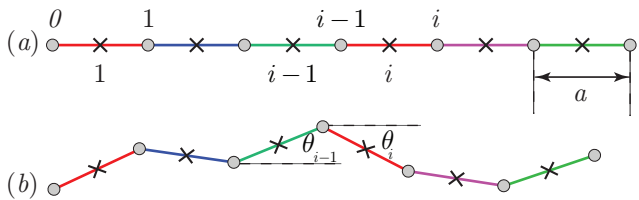}}
\caption{(color online) Schematic of a filament of length $5 \a$ divided into $5$ segments of length $\a$ (a)
in the equilibrium configuration  and (b) in a distorted configuration.
Circles mark lattice nodes (located as positions $i a$), and crosses mark the centers of bonds
located at positions $[i-(1/2)]\a$. The different colors of the bonds indicate different values for the
stretching and bending moduli. The angle of the bonds $i-1$ and $i$ are indicated in (b).  In the limit of
slow changes in $\theta_i$, the slope of the $h(s) \equiv u^{\perp}(s)$ is
constant in bond $i$, and bond angle $i$ is the angle of the line connecting connecting site $i-1$ with
site $i$ for small $\theta_i$.}
\label{FIG:discrete-fil}
\end{figure}

In order to develop a model of crosslinked filaments on a
lattice with a random distribution of stretching and bending
moduli of the sort that we will encounter in our EMTs, we need
first to develop a discretized form of the continuum beam
energy [Eq.~(\ref{EQ:Rod})] with inhomogeneous stretching and
bending moduli. We begin by dividing a filament of length $L$
into $N$ segments (bonds) of length $\a$, labeled $i=1, \cdots
, N$ and terminated by nodes (sites) $i = 0\cdots N$. In
equilibrium in the absence of external forces, the filament is
straight, and node $i$ is at position $s_i=ia$ while that of
the center of bond $i$, which lies between sites nodes $i-1$
and $i$, is at position $s_i -(a/2)$ as shown in
Fig.~\ref{FIG:discrete-fil}. Individual stretching and bending
moduli $\mu_i$ and $\kappa_i$ are associated with bond $i$ as
shown in Fig.~\ref{FIG:discrete-fil}. Individual stretching an
bending moduli $\mu_i$ and $\kappa_i$ are associated with bond
$i$.

The derivation of the discretized stretching energy is
straightforward: Associated with each node $i$  is a
longitudinal displacement $u_i^{||}$ and with each bond $i$ an
energy
\begin{equation}
E_i^s=\frac{1}{2}\frac{\mu_i}{\a}(u_{i+1}^{||} - u_i^{||} )^2 .
\label{EQ:stretchEb}
\end{equation}
The total stretching energy of a filament is the sum of these
bond energies.  The discretized equations of motion arising
from this inhomogeneous discrete model agree with those arising
from a continuum model in the continuum $\a \rightarrow 0$
limit.

The derivation of a discretized bending energy is more subtle.
Consider first a homogeneous model in which $\kappa$ is the
same in each segment.  Here we assign an angle $\theta_i$ to
each bond, and an energy $(1/2) (\kappa/\a^3) (\theta_{i+1} -
\theta_i)^2$ to the node $i$, which lies between bonds $i$ and
$i+1$. This energy is, of course, constructed so that in the
continuum ($\a \rightarrow 0$) limit $(\theta_{i} -
\theta_{i-1})/a \rightarrow d\theta/ds$ and the bending part of
Eq.~(\ref{EQ:Rod}) is retrieved.  This works because the
filament segment between the center of bond $i$ [at position
$s_i a-(\a/2)$] and that of bond $i+1$ [at position $s_i a
+(\a/2)$] is uniform with bending modulus $\kappa$, and as a
result, the energy of that segment is the bond energy given
above. But what happens if the bending moduli in these two
segments are different, i.e., $\kappa_i \neq \kappa_j$?  We
show in App.~\ref{APP:COMP} that the energy of a filament
segment encompassing half of bond $i$ with bending modulus
$\kappa_i$ and half of bond $i+1$ with bending modulus
$\kappa_{i+1}$ is
\begin{equation}
E_{i,i+1}^b = \frac{1}{2}
\frac{\kbar_i}{\a^3}(\theta_{i+1}-\theta_i)^2
\label{EQ:bendingEb}
\end{equation}
where
\begin{equation}
\kbar_i = \frac{2 \kappa_i \kappa_{i+1}}{\kappa_i + \kappa_{i+1}},
\end{equation}
i.e., the two halves of the bending spring connecting bond $i$
to bond $i+1$ add like springs in series.   Note that $\kbar_i$
satisfies the required limits that it reduce to $\kappa_i$ when
$\kappa_{i+1} = \kappa_i$ and that it vanish if either
$\kappa_i$ or $\kappa_{i+1} = 0$.  The total bending energy of
a filament is thus
\begin{equation}
E_{\text{fil}}^b = \frac{1}{2} \sum_{i=1}^N \kbar_i
(\theta_{i+1} - \theta_i)^2 .
\label{EQ:bendingEfil}
\end{equation}
Minimization of this energy gives a series of difference
equations for $\theta_i$.  We show in App.~\ref{APP:COMP} that
the solution to these equations faithfully reproduces
$\theta(s)$ calculated from the continuum equations resulting
from the minimization of the continuum bending energy for the
particular case of $\kappa$'s having one value for $0<s<s_1$
and another value for $s_1<s<L$.  A generalization of this
calculation to more general distributions of $\kappa$ is
straightforward and yields the same results for the discrete
and continuum models.

Ultimately, we are interested in the positions of the nodes,
and we need an expression relating these positions to the bond
angles.  In the ground state, all of the bonds of the filament
are aligned along a common direction specified by a unit
vector, $\ve$, and the ground-state positions are $\vr_i  = s_i
\ve$. Distortions of the filament are described by the
displacement vectors $\vu_i = u_i^{||} \ve + u_i^\perp
\ve^{\perp}$ where $\ve^{\perp}$ is unit vector perpendicular
to $\ve$. As discussed more fully in App.~\ref{APP:COMP},
within the linearized theory we use, the angle that bond $i$
makes with  $\ve$ is then $\theta_i=(u_i^\perp -
u_{i-1}^\perp)/\a$. Thus the bending energy $(1/2) \kbar_i
(\theta_{i+1}-\theta_i)^2\approx (1/2) \kbar_i (2
u_{i}^{\perp}-u_{i+1}^{\perp} - u_{i-1}^{\perp})^2$ couples the
displacements of sites $i-1$, $i$, and $i+1$, and it can be
viewed as an interaction defined on a kind of
next-nearest-neighbor ($NNN$) connecting sites $i-1$ and $i+1$.
This bond, however, only exists if both $NN$ bond $i$ and $i+1$
are occupied.  In what follows, we will refer to the bending
$NNN$ bonds as \emph{phantom} bonds since they do not have an
independent existence. We will also employ an
alternative notation in  which a bond connecting nodes $\ell$
and $\ell'$ on a lattice will be denoted by $\langle \ell,
\ell'\rangle$ and the angle that bond makes with the horizontal
axis by $\theta_{\langle \ell, \ell' \rangle}$.  The $NNN$
bending energy is then $(1/2) \kappa (\theta_{\ell,
\ell',\ell''})^2$, where $\theta_{\ell,
\ell',\ell"}=\theta_{\langle \ell, \ell'
\rangle}-\theta_{\langle \ell',\ell''\rangle}$ with the
understanding that sties $\ell$, $\ell'$, and $\ell''$ all lie
on a single filament.

\subsection{Triangular lattice of filamentous polymers}

\begin{figure}
    \centering
     {\includegraphics[width=.3\textwidth]{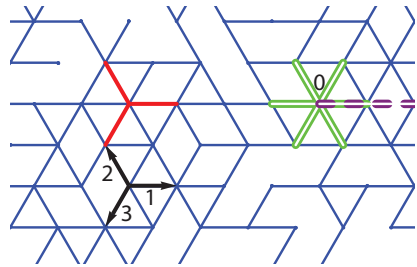}}
        \caption{(color online) Filamentous triangular lattice with bonds randomly occupied
        with probability $\Prob$.  The unit vectors $\uvepara_1$, $\uvepara_2$, $\uvepara_3$
        are marked by ``1, 2, 3'', the 3 stretch energy vectors $\vB^{s}_{n}$ are marked by
        the 3 red single lines, and the 3 bending energy vectors $\vB^{b}_{n}$ are
        marked by the 3 green double lines.  The purple dashed double line marks the
        bending vector $\vB^{b}_{4}$ if the origin is marked by $0$.
        }
        \label{FIG:TrigLatt}
\end{figure}

To create a network of crosslinked semiflexible polymers we
randomly remove bonds on a triangular lattice.  Polymers
correspond to lines of connected, occupied colinear bonds, and
crosslinks correspond to sites at which two or three polymers
cross. Each bond in the lattice can be assigned one of the
three directions designated by the unit vectors $\uvepara_n$
shown in Fig.~\ref{FIG:TrigLatt}.  All of the bonds in a given
filament are aligned along one of these directions and the
filament itself is directed. Sites on the lattice are labeled
by a two-component index $\ell = (l_1, l_2)$, and their
equilibrium positions are  $\vr_{\ell} = \a(l_1 \uvepara_1 +
l_2 \uvepara_2)$.  We adopt the convention that $NN$ bonds
$\langle \ell,\ell'\rangle$ connect sites with equilibrium
positions $\vr_\ell$ and $\vr_{\ell'} = \vr_\ell + \a
\uvepara_n$ for one of the directions $\uvepara_n$. Upon
distortion, the position of site $\ell$ changes to $\vR_{\ell}
= \vr_{\ell} + \vu_{\ell}$, where $\vu_{\ell}$ is the
displacement vector of site $\ell$.  We define all bond angles
to be zero in the undistorted lattice.  In the distorted
lattice, the angle of bond $\langle \ell, \ell' \rangle$
becomes $\theta_{\langle \ell, \ell'\rangle} \approx \vu_{\ell,
\ell'}\cdot \uvr_{\ell,\ell'}^\perp/\a$, where $\vu_{\ell,
\ell'} = \vu_{\ell'} - \vu_{\ell}$ and
$\uvr_{\ell,\ell'}^\perp$ is the unit vector perpendicular to
the bond direction along $\vr_{\ell'} - \vr_{\ell}$ and is
equal to one of the unit vectors $\uveperp_n$ perpendicular to
$\ve_n$. We now assign stretching energies to each bond and
bending energies to each phantom $NNN$ bond along a lattice
direction in accordance with the discretized energy of an
individual filament [Eqs.~(\ref{EQ:stretchEb}) and
(\ref{EQ:bendingEb})] to obtain the harmonic energy on a
diluted lattice
\begin{subequations}\label{EQ:Hami}
\begin{eqnarray}
		E &=& E_{s}+E_{b} \\
    E_{s} &=& \frac{1}{2} \frac{\mu}{\a} \sum_{\langle \ell,\ell'\rangle}
    g_{\ell, \ell'} (\vu_{\ell\ell'}\cdot \uvr_{\ell\ell'})^2 , \label{EQ:ENST}\\
    E_{b} &=& \frac{1}{2} \frac{\kappa}{\a}
    \sum_{\ell',n} g_{\ell,\ell'} g_{\ell',\ell''}( \theta_{\ell\ell'\ell''})^2 \\
    &=& \frac{1}{2} \frac{\kappa}{\a^3} \sum_{\langle \ell,\ell',\ell''\rangle}
    g_{\ell,\ell'} g_{\ell',\ell''}\big\lbrack (\vu_{\ell\ell'}-\vu_{\ell'\ell''})
    \cdot \uvr_{\ell,\ell'}^\perp \big\rbrack^2, \nonumber \label{EQ:ENBE}
\end{eqnarray}
\end{subequations}
where $g_{\ell,\ell'} = 1$ if the bond
$\langle\ell,\ell'\rangle$ is occupied and $g_{\ell,\ell'} = 0$
if it is not. This is the model was introduced in
references \cite{Wilhelm2003} and \cite{Head2005} in their
study of the Mikado model. A version of this model in which
there is a bond-angle energy between all pairs of bonds sharing
a common site rather than only between pairs of parallel bonds
was introduced earlier in reference \cite{FengLob1984}.

When $\Prob=1$, all bonds are occupied and $E$ becomes
homogeneous.  In this limit, the long-wavelength elastic energy
reduces to the elastic energy of an isotropic $2d$ medium,
\begin{eqnarray}\label{EQ:CONT}
    E=\int d^2 x \left[\frac{\bar{\lambda}}{2} \big(\textrm{Tr}\underline{\underline{u}}\big)^2 +
    \bar{\mu} \textrm{Tr}\underline{\underline{u}}^2 \right],
\end{eqnarray}
where $\underline{\underline{u}}$ is the linearized symmetric
Cauchy strain tensor with Cartesian components $u_{ij}$,
$\bar{\lambda}$ and $\bar{\mu}$ are the Lam\'{e} coefficients,
$\bar{\lambda}=\bar{\mu}=(\sqrt{3}/4)(\s /\a)$, which depend
only on $\mu$ and not on $\kappa$. $\bar{\mu}$  is the
macroscopic shear modulus. The bending constant $\b$ only
appears in the higher-order gradients of the the displacement
vector. Upon dilution, each of the bonds is present with a
probability $\Prob$, and the resulting lattice corresponds to a
random network of semiflexible filaments of finite random
lengths $L$, whose average as a function of $\Prob$ is $\langle
L \rangle = \a/(1-\Prob)$~\cite{BroederszMac2011}. It is a
straightforward exercise to show that the average distance
$l_c$ between crosslinks (i.e. nodes with at which two or more
filaments cross) differs by at most a factor of $2$ from $\a$,
and we will use treat them as the same quantity in what
follows. In EMT, $\mu$ is replaced in diluted samples by its
effective medium value $\sm$, and the macroscopic EMT shear
modulus of these samples is
\begin{equation}
G = (\sqrt{3}/4)(\sm /\a) .
\end{equation}
In the undiluted limit the shear modulus is $G_0
=(\sqrt{3}/4)(\mu /\a)$, and $G/G_0 = \mu_m/\mu$. Because our
calculations are centered on the evaluation of $\mu_m$ rather
than $G$, we will in what follows use $\mu_m$ as a proxy for
$G$, reminding the reader where appropriate of this simple
relation between the effective medium parameter and $G$.

\section{Effective Medium Theory \label{SEC:CPA}}

We study the elasticity of our network using an
effective-medium approximation~\cite{Soven1969,Elliott1974} in
which the random inhomogeneous system is replaced with an
effective homogeneous one constructed so that the average
scattering  from a bond (or chosen set of bonds) with the
probability distribution of the original random lattice
vanishes. In more technical terms, the effective medium is
chosen so that average $T$-matrix associated with the bond
vanishes. This approximation has been shown to be a powerful
tool for the calculation of properties of random systems, from
the electronic structure of
alloys~\cite{Soven1969,Kirkpatrick1970} to the elasticity of
random networks~\cite{Feng1985,Mao2010}.

Our elastic energy is a bilinear form in the $2N$-dimensional
displacement vector $\vu$ determined by the $2N \times 2N$
dynamical matrix $\DMT$, where $N$ is the number of sites in
the lattice. We will represent these two quantities in both the
lattice basis and the wavenumber basis where the components of
$\vu$ are, respectively, the $2$-dimensional vectors $\vu_\ell$
and $\vu_\vq$ for each of the $N$ lattice positions $\ell$ or
wavenumbers $\vq$ and the components of $\DMT$ are respectively
the $2\times2$ matrices $\DMT_{\ell,\ell'}$ and
$\DMT_{\vq,\vq'}$ for each pair $(\ell,\ell')$ or $(\vq,\vq')$.
We use the convention in which arbitrary vectors $\vv$ or
matrices $\MM$ in the two bases are related via
\begin{align}\label{EQ:FTDefi}
    \vv_{\vq} & =\sum_{\ell} \vv_{\ell} e^{-i \vq \cdot \vr_{\ell}} , \qquad
    \vv_{\ell} = \frac{1}{N} \sum_{\vq} \vv_{\vq} e^{i \vq \cdot \vr_{\ell}} ,\\
    \DMT_{\vq,\vq'} & =\sum_{\ell,\ell'} e^{-i \vq \cdot \vr_{\ell}} \DMT_{\ell,\ell'}
    e^{i \vq' \cdot \vr_{\ell'}}, \nonumber\\
    \DMT_{\ell,\ell'} & =\frac{1}{N^2} \sum_{\vq,\vq'} e^{i \vq \cdot \vr_{\ell}} \DMT_{\vq,\vq'}
    e^{-i \vq' \cdot \vr_{\ell'}} .
\end{align}
The elastic energy is thus
\begin{align}
&E= \frac{1}{2} \vu\cdot\DMT\cdot\vu \nonumber \\
& =\frac{1}{2}\sum_{\ell,\ell'}
\vu_{\ell}\cdot\DMT_{\ell,\ell'}\cdot \vu_{\ell'}= \frac{1}{2
N^2}\sum_{\vq,\vq'}\vu_{-\vq} \cdot \DMT_{\vq,\vq'}\cdot
\vu_{\vq'} ,
\end{align}
where here and in the following the ``dot'' signifies the
multiplication of a matrix and a vector or of two matrices. The
zero-frequency phonon Green's function (which is a $2N\times2N$
matrix) is minus the inverse of the dynamical matrix:
\begin{equation}
\GF = -\DMT^{-1} .
\end{equation}

In EMT, the inhomogeneous and random dynamical matrix $\DMT$ is
replaced by a homogeneous, translationally invariant one $\DM$,
with $\DM_{\ell,\ell'} = \DM_{\ell - \ell'}$ and
\begin{equation}
\DM_{\vq,\vq'} = N \delta_{\vq,\vq'} \DM_\vq ,
\end{equation}
along with a perturbation matrix $\PPV$, which we will specify
in detail shortly:
\begin{equation}
\DMT=\DM + \PPV = - (\GF^{(m)})^{-1} + \PPV = -\GF^{-1} ,
\end{equation}
where the superscript $(m)$ stands for \lq\lq effective
medium\rq\rq. The full Green's function can thus be expressed
as
\begin{equation}
\GF = ((\GF^{(m)})^{-1} - \PPV)^{-1} = \GF^{(m)}+\GF^{(m)}\cdot\TM\cdot \GF^{(m)} ,
\end{equation}
where $\TM$ is the $T$-matrix describing the scattering
resulting from $\PPV$:
\begin{eqnarray}\label{EQ:Texp}
\TM & = & \PPV\cdot(\IM - \PPV \cdot\GF^{(m)})^{-1} =
(\IM- \PPV \cdot \GF^{(m)}))^{-1}\cdot \PPV \nonumber \\
&= &\PPV + \PPV \cdot \GF^{(m)} \cdot \PPV + \cdots .
\end{eqnarray}
This expresses the $T$-matrix in general form.  Our next step
is to specify both $\DM$ and $\PPV$.

\begin{figure}%[t]
    \centering
     \includegraphics[width=.45\textwidth]{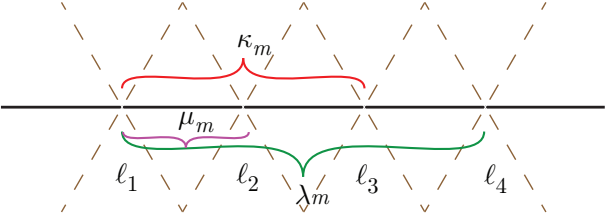}
        \caption{(Color online) Positions of sites $\ell_1,\ell_2,\ell_3,\ell_4$,
        and interactions in the effective medium, including NN stretching term of rigidity $\mu_m$,
        NNN bending term of rigidity $\kappa_m$, and third neighbor effective coupling term of rigidity $\lambda_m$.
        }
\label{FIG:EMT}
\end{figure}

We begin with $\DM$.  Normally, the effective-medium elastic
energy would simply be the random one of Eq.~(\ref{EQ:Hami})
with $\mu$ and $\kappa$ replaced by their respective
effective-medium values $\mu_m$ and $\kappa_m$ and
$g_{\ell,\ell'}$ replaced by one.  It turns out, however, as we
will shortly demonstrate, that the effective-medium equations,
determined by setting the average $T$-matrix equal to zero,
consists of three independent equations whose solutions
requires three independent parameters.  If the above simple
procedure for constructing the effective-medium energy is
followed, there are only two parameters, $\mu_m$ and
$\kappa_m$, and to solve the EMT equations, it is necessary to
introduce a new term to this energy with a new parameter, which
we denote by $\lm$. This additional energy, whose form is
dictated, as we shall see, by the EMT equations, couples angles
on neighboring $NNN$ phantom bonds: 
\begin{equation}\label{EQ:Eeff_bb}
E_{bb}(\lambda_m) = \frac{\lambda_m}{\a^3} \sum_{\ell_2}
\theta_{\ell_1,\ell_2,\ell_3} \theta_{\ell_2,\ell_3,\ell_4} ,
\end{equation}
where it is understood that the sites
$\ell_,\ell_2,\ell_3,\ell_4$ are sequential sites along a
filament as shown in Fig.~\ref{FIG:EMT}. The total
effective-medium energy is thus
\begin{equation}
E^{(m)} (\mu_m,\kappa_m,\lambda_m)= E_s (\mu_m) + E_b(\kappa_m)
+E_{bb}(\lambda_m) ,
\end{equation}
and its associated dynamical matrix is
\begin{align}\label{EQ:DMCons}
    &\DM_{\vq}(\sm,\bm)
    = \frac{\sm}{\a} \sum_{n=1}^{3} \vB^{s}_{n,\vq} \vB^{s}_{n,-\vq} \nonumber\\
   & + \frac{\bm}{\a^3} \sum_{n=1}^{3} \vB^{b}_{n,\vq} \vB^{b}_{n,-\vq}
 + \frac{\lm}{\a^3} \sum_{m=1}^{3} 2\cos(\vq\cdot\uvepara_m) \vB^{b}_{m,\vq} \vB^{b}_{m,-\vq}
\end{align}
where
\begin{subequations}
\begin{eqnarray}
    \vB^{s}_{n,\vq} &=& (1-e^{-i\vq\cdot\uvepara_n}) \uvepara_n , \label{EQ:vBs}\\
    \vB^{b}_{n,\vq} &=& 2[1-\cos(\vq\cdot\uvepara_n)] \uveperp_n  ,  \label{EQ:vBb}
\end{eqnarray}
\end{subequations}
are two-dimensional vectors and where a simplified notation is
used in which two of these vectors in a row denotes a direct
product creating a $2\times2$ matrix.

The perturbation $\PPV$ arises from the removal of a single
bond, whose  endpoints, $\ltwo$ and $\lthree$, we take to be
contiguous sites along a filament parallel to the $\uvepara_1$
axis  with $\ltwo$ located at the origin and $\lthree$ at
position $\uvepara_1$. If there is no bending energy (i.e.,
$\kappa = 0$), the energy of this bond relative to the
effective medium is thus
\begin{equation}
E_V^s= \frac{1}{2} \frac{\mu_s - \mu_m}{\a}[(\vu_{\ltwo}-\vu_{\lthree})\cdot \uvepara_1]^2 ,
\end{equation}
where $\mu_s = \mu g_{\ltwo,\lthree}$ so that its probability
distribution is
\begin{equation}
    P(\ss) = \Prob \delta(\ss-\s) + (1-\Prob) \delta(\ss)  .
\end{equation}
This bond stretching energy defines $\PPV^s$:
\begin{equation}
\PPV_{\vq,\vq'}^s=\a^{-1}(\ss - \sm) \vB^{s}_{1,\vq} \vB^{s}_{1,-\vq'}.
\end{equation}
Note that $\PPV_{\vq,\vq'}^s$ factorizes into a product of a
term depending only on $\vq$ and a term depending only on
$\vq'$. This is a property, shared by the other contributions
to $\PPV$, that, as we shall see, makes the calculation of the
$T$-matrix from Eq.~(\ref{EQ:Texp}) tractable.

Replacing bond $\langle \ltwo,\lthree \rangle$ changes the
bending as well as the stretching modulus of that bond.  As
discussed in Sec.~\ref{SEC:MODEL} and App.~\ref{APP:COMP}, this
leads to a change in the bending constant of the $NNN$ bonds
$\langle \lone,\ltwo,\lthree \rangle$ and $\langle
\ltwo,\lthree,\lfour \rangle$ that share the replaced bond
$\langle \ltwo,\lthree \rangle$ along a filament from
$\kappa_m$ to
\begin{eqnarray}\label{EQ:Vbeta}
    \btwo=2\Big(\frac{1}{\b_s}+\frac{1}{\b_m}\Big)^{-1},
\end{eqnarray}
where $\b_s \equiv \kappa g_{\ltwo,\lthree}$ equals zero if the
bond $\langle \ltwo\lthree \rangle$ is vacant and $\b$ if it is
occupied. The probability distribution for $\b_s$ is thus
\begin{equation}
P(\b_s) = p\delta (\bs - \b) + (1-p) \delta (\bs) ,
\end{equation}
and the joint probability distribution for both $\ss$ and $\bs$
is
\begin{equation}
P(\ss,\bs) = \Prob \delta(\ss-\s)\delta(\bs-\b) + (1-\Prob) \delta(\ss)\delta(\bs) .
\end{equation}

If $\langle \ltwo\lthree \rangle$ is occupied $\b_c = 2 \b
\b_m/(\b + \b_m)$ is a nonlinear function of $\b$ and $\b_m$.
These considerations determine the bending contribution to
$E_V$,
\begin{align}\label{EQ:EV_bb}
E_V^b  = & \frac{1}{2}\frac{\btwo(\bs)-\bm}{\a^3}
\left[(\theta_{\lone,\ltwo,\lthree})^2
+ (\theta_{\ltwo,\lthree,\lfour})^2\right] \nonumber\\
= & \frac{1}{2}\frac{\btwo(\bs)-\bm}{\a^3}
	\Big\lbrace \lbrack (2\vu_{\ltwo} - \vu_{\lthree} - \vu_{\lone})
	\cdot \uveperp_{1} \rbrack^2 \nonumber\\
	&\quad+
	\lbrack (2\vu_{\lthree} - \vu_{\lfour} - \vu_{\ltwo})\cdot \uveperp_{1} \rbrack^2 \Big\rbrace ,
\end{align}
and the bending contribution to $\PPV$:
\begin{eqnarray}
\PPV^b_{\vq,\vq'} & = &
  \a^{-3}(\btwo(\bs)-\bm)  \vB^{b}_{1,\vq} \vB^{b}_{1,-\vq'} \nonumber\\
    && + \a^{-3}(\btwo(\bs)-\bm)  \vB^{b}_{4,\vq} \vB^{b}_{4,-\vq'} ,
\end{eqnarray}
where the vectors $\vB^{b}_{1,\vq}$ [Eq.~(\ref{EQ:vBb})] and
$\vB^{b}_{4,\vq}\equiv e^{-i\vq\cdot\uvepara_1}\vB^{b}_{1,\vq}$
represent the bending of the bond pair connecting sites
$\lone,\ltwo,\lthree$ and $\ltwo,\lthree,\lfour$, respectively.
Finally, the original energy had no term corresponding the
coupling between $\theta_{\lone, \ltwo, \lthree}$ and
$\theta_{\ltwo,\lthree,\lfour}$ that appears in the effective
medium energy [Eq.~\ref{EQ:Eeff_bb}], so that replacement of
the bond $\langle\ltwo,\lthree\rangle$ with its form in the
original energy removes the energy associated with that bond in
$E^{(m)}$ and creates the contribution
\begin{equation}
\PPV_{\vq,\vq'}^{bb} = - \frac{\lambda_{m}}{\a^3}[\vB^{b}_{1,\vq}
\vB^{b}_{4,-\vq'} + \vB^{b}_{4,\vq}\vB^{b}_{1,-\vq'} ]
\end{equation}
to $\PPV$.  The complete $\PPV$ is thus $\PPV =
\PPV^s+\PPV^b+\PPV^{bb}$, which can conveniently be expressed
as
\begin{equation}
	\PPV_{\vq,\vq'}(\ss,\bs)  = \sum_{\alpha,\beta} \tPPV^{\alpha\beta}(\ss,\bs)
	\vB_{\vq}^{\alpha}  \vB_{-\vq}^{\beta} ,
\end{equation}
where $\alpha=\{(s,1), (b,1), (b,4)\}$ labels the three vectors
$\{\vB^{s}_{1,\vq},\vB^{b}_{1,\vq},\vB^{b}_{4,\vq}\}$. The
scattering potential in this basis is
\begin{widetext}
\begin{eqnarray}\label{EQ:tPPV}
    \tPPV(\ss,\bs,\lm) =\left( \! \begin{array}{ccc}
        (\ss-\sm)/\a & 0 & 0 \\
        0 & (\btwo(\bs)-\bm)/\a^3 & -\lm/\a^3 \\
        0 & -\lm/\a^3 & (\btwo(\bs)-\bm)/\a^3 \\
    \end{array} \! \right),
\end{eqnarray}
\end{widetext}

We are now in a position to calculate the $T$-matrix.  Consider
first the first non-trivial term in its series expansion
[Eq.~(\ref{EQ:Texp})]:
\begin{align}
\PPV \cdot \GF \cdot \PPV  \rightarrow &
\frac{1}{N}\sum_{\vq_1}\! \PPV_{\vq,\vq_1}\!\!\cdot\!
    \GF_{\vq_1}^{(m)}\!\!\cdot\!\PPV_{\vq_1,\vq'}
    \nonumber\\
    =& \sum_{\alpha,\beta,\alpha',\beta'} \vB_{\vq}^{\alpha}
    \tPPV^{\alpha\beta} \cdot (\tGF^{(m)})^{\beta\alpha'}
     \cdot \tPPV^{\alpha'\beta'} \vB_{-\vq'}^{\beta'}
\end{align}
where $\tGF$ is defined as
\begin{align}\label{EQ:DEFtG}
	(\tGF^{(m)})^{\beta,\alpha'} \equiv \frac{1}{N} \sum_{\vq_1}\!
     \vB_{-\vq_1}^{\beta}\cdot\!\GF^{(m)}_{\vq_1}\!\!\cdot\!\vB_{\vq_1}^{\alpha'} .
\end{align}
It is clear that subsequent terms in the Taylor series for
$\TM$ decompose in a similar way and that
\begin{equation}
\TM_{\vq,\vq'} =\sum_{\alpha,\beta} \vB_{\vq}^{\alpha}
    \tTM^{\alpha\beta}  \vB_{-\vq'}^{\beta}
\end{equation}
where the $3\times 3$ matrix $\tTM$ satisfies.
\begin{equation}\label{EQ:tTM}
\tTM = \tPPV(\tIM-\tGM \tPPV)^{-1} = (\tIM-\tPPV \tGM)^{-1} \tPPV
= (\tPPV^{-1} - \tGF)^{-1} ,
\end{equation}
where $\tPPV \tGM$ signifies a matrix product.

There are now a couple of points that must be attended to
before we present the details or our calculation. First, we
show in App.~\ref{APP:DM} that $\tGM$ is a symmetric matrix
whose $12$ and $13$ components vanish and whose $22$ and $33$
components are equal whether or not $\lm$ is zero. Importantly,
the $23$ component of $\tGM$ is nonzero even if $\lm$ is zero.
Thus, $\tGM$ has the same structure as $\tPPV$:
\begin{eqnarray}\label{EQ:tGF}
    \tGM =\left( \begin{array}{ccc}
        G^{(m)}_1 & 0 & 0 \\
        0 & G^{(m)}_2 & G^{(m)}_3 \\
        0 & G^{(m)}_3 & G^{(m)}_2 \\
    \end{array} \right),
\end{eqnarray}
where $G^{(m)}_1=(\tGM)^{11}$, $G^{(m)}_2=(\tGM)^{22}$, and
$G^{(m)}_3=(\tGM)^{23}$. This implies from Eq.~(\ref{EQ:tTM})
that $\tTM$ also has the same structure as $\tGM$ with three
independent components ($\tT^{11}, \tT^{22}=\tT^{33}$, and
$\tT^{23}$) even if $\lm=0$.  Thus, the EMT equation
\begin{equation}\label{EQ:TM-0}
\langle \tTM \rangle  = \Prob \tTM(\ss=\s,\bs=\b) +(1-\Prob) \tTM(\ss=0,\bs=0) =0 ,
\end{equation}
reduces to three independent equations whose solution requires
three independent parameters.  The addition of the energy
$E_{bb}^{(m)}$ [Eq.~(\ref{EQ:Eeff_bb})] adds the needed third
parameter, $\lm$, to $\sm$ and $\bm$ and gives $\tPPV$ the same
structure as $\tTM$ and and $\tGF$.

To solve Eq.~(\ref{EQ:TM-0}), we first write it as
\begin{align}\label{EQ:TM-p}
& p\tPPV(\mu,\kappa)[\tIM - \tGM \tPPV(\mu,\kappa)]^{-1} \nonumber\\
& + (1-p) [\tIM - \tPPV(0,0)\tGM ]^{-1} \tPPV(0,0)  = 0 ,
\end{align}
where we used both forms of Eq.~(\ref{EQ:tTM}).  Multiplying
this equation on the left by $[\tIM - \tGM \tPPV(0,0)]^{-1}$
and on the right by $[\tIM - \tPPV(\mu,\kappa)\tGM ]^{-1}$, we
obtain
\begin{equation}
\Prob \tPPV(\s,\b) + (1-\Prob) \tPPV(0,0) - \tPPV(0,0)\tGM\tPPV(\s,\b) =0,
\label{EQ:EMT-2}
\end{equation}
which has the advantage that it contains no inverse matrices.
At this point, it is convenient to introduce the reduced Green
function
\begin{equation}
  \tilde{H}(\bmr,\lmr) \equiv -\frac{\sm}{\a} \tGM(\sm,\bm,\lm) .
\end{equation}
\vspace{12 pt}
From the definition of $\tGM$ it is straightforward to see that
$\tilde{H}$ only depends on the ratios $\bmr\equiv\bm/\sm$ and
$\lmr\equiv\lm/(\sm \a^2)$.  Clearly $\tilde{H}$ has the same
structure as $\tGM$ with $H_\sigma=-(\sm/\a) G^{(m)}_\sigma,$
for $\sigma=1,2,3$. With these definitions, the $11$ component
of  equation Eq.~(\ref{EQ:EMT-2}) is
\begin{equation}\label{EQ:AsymEqOne}
    \sm=\s\frac{\Prob-H_{1}(\bmr,\lmr)}{1-H_{1}(\bmr,\lmr)}.
\end{equation}
and the $22$ and $23$ components are, respectively,
\begin{widetext}
\begin{align}
    &2\Big(\frac{1}{\br}+\frac{1}{\bmr}\Big)^{-1}
    \Big(\Prob-\frac{1}{2}\big(1+\frac{\bmr}{\br}\big) -
    \bmr H_{2}-\lmr H_{3} \Big) +(\bmr^2+\lmr^2)H_2 + 2\bmr\lmr H_{3} =0 \label{EQ:AsymEqTwo},\\
    & -\lmr - 2\Big(\frac{1}{\br}+\frac{1}{\bmr}\Big)^{-1}
    \big(\lmr H_{2}+\bmr H_{3} \big) +2\bmr\lmr H_2 + (\bmr^2+\lmr^2)H_{3} =0 \label{EQ:AsymEqThe},
\end{align}
\end{widetext}
where $b=\kappa/(\mu_m a^2)$. Thus we have 3 unknowns
$\{\sm,\bmr,\lmr\}$ (or equivalently, $\{\sm,\bm,\lm\}$) and
$3$ equations Eq.~(\ref{EQ:AsymEqOne}),
Eq.~(\ref{EQ:AsymEqTwo}), and Eq.~(\ref{EQ:AsymEqThe}). These
are our exact EMT equations.

\subsection{Scaling Solutions near $\PCF$}

Here we solve the EMT self-consistency equations,
Eqs.~(\ref{EQ:AsymEqOne}) to (\ref{EQ:AsymEqThe}) near $\PCF$
at small $\b$.  When $\b=0$ the problem reduces to that of a
central-force rigidity percolation \cite{Feng1984} with zeroth
order solutions $\bm^0=0$, $\lm^0=0$ , and
\begin{equation}
\sm^{(0)}=\s\frac{\Prob-\PCF}{1-\PCF} ,
\end{equation}
where $\PCF = H_1(0,0)=2/3$ which can also be obtained via symmetry arguments~\cite{Feng1985}. As $\b$ increases from zero, $\sm$
increases, $\bmr$ and $\lmr$ become nonzero, and the rigidity
threshold jumps to a lower value $\Pb$ as shown in
Fig.~\ref{FIG:CPASOLUK}(b). For small $\b$, we have $\b/(\s
\a^2)\ll 1$,  we can assume that $\bmr,\lmr\ll 1$ (which we
will verify later), and we find that to the leading order the
three Eqs.~(\ref{EQ:AsymEqOne}), (\ref{EQ:AsymEqTwo}), and
(\ref{EQ:AsymEqThe}) become
\begin{subequations}\label{EQ:smbmlm}
\begin{eqnarray}
    \sm &\simeq& \s\frac{\Prob-\PCF-H_{1,1}(0,0)\bm/(\sm \a^2)}{1-\PCF} ,\\
    \bm &\simeq& \b(2\Prob-1) ,\\
    \lm &\simeq& \b H_{3}(0,0) \frac{\b}{\sm \a^2} \frac{1-\Prob}{\Prob} (2\Prob-1)^2,
\end{eqnarray}
\end{subequations}
where $H_{1,1}(0,0) = \partial H_{1} /\partial \bmr
\vert_{\bmr=0,\lmr=0}\simeq -2.413$ and $H_3(0,0) = 1.520$. For
convenience we define $\mathcal{A}\equiv -H_{1,1}(0,0)$ and
$\mathcal{B}\equiv H_{3}(0,0)$.  From these relations, we find
that at $\Prob=\PCF$
\begin{subequations}
\begin{eqnarray}
    \sm &\sim& \b^{1/2} ,\\
    \bm &\sim& \b ,\\
    \lm &\sim& \b^{3/2},
\end{eqnarray}
\end{subequations}
indicating that $\sm\gg\bm\gg\lm$ and thus $\bmr,\lmr\ll 1$ as
we assumed. Using these relations, together with the fact that
as $\b\to 0$, $\sm\to  \Prob-\PCF$, we solve
Eqs.~(\ref{EQ:smbmlm}) to obtain
\begin{subequations}
\begin{eqnarray}\label{EQ:Scal}
    \sm &=& \s\vert\Delta\Prob\vert^{t_1}
     g_{1,\pm}\Big(\frac{\b}{\a^2 \s\vert\Delta\Prob\vert^{\phi}}\Big), \\
    \bm &=& \s a^2 \vert\Delta\Prob\vert^{t_2}
     g_{2,\pm}\Big(\frac{\b}{\a^2 \s\vert\Delta\Prob\vert^{\phi}}\Big) ,\\
    \lm &=& \s a^2\vert\Delta\Prob\vert^{t_3}
     g_{3,\pm}\Big(\frac{\b}{\a^2 \s\vert\Delta\Prob\vert^{\phi}}\Big),
\end{eqnarray}
\end{subequations}
where
\begin{subequations}
\begin{eqnarray}\label{EQ:Expo}
    \phi&=&2 ,\\
    t_1 &=& 1, \\
    t_2&=&2 ,\\
    t_3&=&3 ,
\end{eqnarray}
\end{subequations}
and
\begin{subequations}
\begin{eqnarray}\label{EQ:ScalFunc}
	g_{1,\pm}(x) & \simeq &
		\frac{3}{2}\left( \pm 1+\sqrt{1-\frac{4\mathcal{A}}{9}x} \right) ,\\
	g_{2,\pm}(x) & \simeq &
		\frac{1}{3}x ,\\
	g_{3,\pm}(x) & \simeq &
		\frac{\mathcal{B}}{27} \left( \pm 1+\sqrt{1-\frac{4\mathcal{A}}{9}x} \right)^{-1} x^2 ,
\end{eqnarray}
\end{subequations}
These scaling relations are analogous to that found in random
resistor networks with two different types of
resistors~\cite{Straley1976}, and central force spring networks
with strong and weak springs~\cite{WyartMah2008}.

Thus, the EMT modulus in the vicinity of $\PCF$ is
\begin{eqnarray}\label{EQ:gOne}
    \sm &=& \s \vert\Delta\Prob\vert
    \frac{3}{2}\Bigg( \pm 1+\sqrt{1
    	-\frac{4\mathcal{A}}{9}\frac{\b}{\a^2 \s\vert\Delta\Prob\vert^{\phi}}} \Bigg) \\
    &\simeq&
        \begin{cases}
            \frac{\sqrt{\mathcal{A}}}{\a} \s^{1/2}\b^{1/2}
            & \text{if
        $\frac{\b}{\a^2 \s\vert\Delta\Prob\vert^{\phi}} \gg 1$,}
        \\
        3 \s\vert\Delta\Prob\vert
         & \text{if $\frac{\b}{\a^2 \s\vert\Delta\Prob\vert^{\phi}} \ll 1$ and $\Delta \Prob>0$,}\\
        \frac{\mathcal{A}}{3\a^2} \frac{\b}{\vert\Delta\Prob\vert} &
        \text{if $\frac{\b}{\a^2 \s\vert\Delta\Prob\vert^{\phi}} \ll 1$ and $\Delta \Prob<0$.}
        \end{cases} \nonumber .
\end{eqnarray}
These crossover regimes correspond exactly to those found in
Ref.~\cite{WyartMah2008} using known behavior of the density of
states and mode structure of systems near the CF isostatic
limit and general scaling arguments.

\subsection{Solutions near $\Pb$}

Equations~(\ref{EQ:AsymEqOne}) to (\ref{EQ:AsymEqThe}) can also
be used to solve for the asymptotics near the rigidity
threshold $\Pb$.  In particular, because $\lmr,\bmr$ converge
to constants that are much smaller than unity and independent
of $\b$ near $\Pb$, the asymptotic solution near $\Pb$ in this
section are not limited to small $\b$.

Firstly, we solve for the value of the rigidity threshold $\Pb$
for the case of $\b>0$ using these EMT equations.  At $\Pb$, we
have $\sm=0,\bm=0,\lm=0$ and as a result $\br\to\infty$.
The ratios $\bmr$ and $\lmr$ are, however, not zero,
and we solve for them. So the equations that determine
$\Pb,\bmr=b_b,\lmr=l_b$ are
\begin{eqnarray}
    \Pb - H_{1}(b_{b},l_{b}) &=& 0 , \nonumber\\
    2b_{b}\Big(\Pb -\frac{1}{2}\Big) + (-b_{b}^2 + l_{b}^2) H_{2}(b_{b},l_{b})  &=& 0 , \nonumber\\
    -l_{b} + (-b_{b}^2 +l_{b}^2) H_{3}(b_{b},l_{b})&=& 0 ,
\end{eqnarray}
where $b_b$ and $l_b$ are the value of $\bmr$ and $\lmr$ at
$\Pb$. This set of equations is independent of $\b$. Numerical
solutions to these equations are given by
\begin{eqnarray}
    \Pb &\simeq& 0.5584 , \nonumber\\
    b_b &\simeq& 0.06355 , \nonumber\\
    l_b &\simeq& 0.004235 ,
\end{eqnarray}
which agrees with the results we obtained by solving the EMT
equations numerically.

Secondly, we solve for the asymptotic behaviors near $\Pb$.  To
achieve this, we suppose $\Prob=\Pb+\deltap$, and to first
order we have
\begin{eqnarray}\label{EQ:Expansion}
	\sm=0+\dsm \nonumber\\
	\bmr=b_b+\dbm \nonumber\\
	\lmr=l_b+\dlm .
\end{eqnarray}
We put these expansions back into
Eqs.~(\ref{EQ:AsymEqOne},\ref{EQ:AsymEqTwo},\ref{EQ:AsymEqThe})
we get the first order perturbation equations
\begin{eqnarray}
	&\dsm = \frac{\s}{1-\Pb} (\deltap - A_1 \dbm - A_2 \dlm ) , \nonumber\\
	&(2\Pb-1-2b_b H_{2,0}) \dbm = -2b_b \deltap + 2\frac{\dsm \a^2}{\b}\Pb b_b^2 , \nonumber\\
	&\dlm = -2b_b H_{3,0} \dbm . \label{EQ:eqThree}
\end{eqnarray}
where $A_1\equiv H_{1,1}(b_b,l_b)\simeq -1.371$ and $A_2\equiv
H_{1,2}(b_b,l_b)\simeq 1.474$.  In deriving these equations we
used the fact that $l_b\ll b_b \ll 1$ and
$\frac{\bmr}{\br}=\frac{\bm}{\b}\ll 1$ near $\Pb$. Thus we
arrive at the asymptotic solution of the effective medium
stretching stiffness
\begin{eqnarray}\label{EQ:umAsym}
\sm = \s \b \frac{c_2 \deltap}{\b+c_1\a^2 \s}
\end{eqnarray}
where
\begin{eqnarray}
	c_1 &=& \frac{A_1 - 2b_b H_{3,0} A_2}{2\Pb-1-2b_b H_{2,0}} \frac{2\Pb b_b^2}{1-\Pb} \nonumber\\
	c_2 &=& \frac{1}{1-\Pb} \left( 1+2b_b \frac{A_1 - 2b_b H_{3,0} A_2}{2\Pb-1-2b_b H_{2,0}} \right)
\end{eqnarray}
are constants determined by the architecture of the lattice and
are independent of $\Prob$ or $\b/(\s \a^2)$.  In the case of
triangular lattice we have $c_1=0.1018$ and $c_2=5.132$.

\section{{Numerical Results} \label{SEC:Results}}
\begin{figure}[h]
    \centering
     \subfigure[]{\includegraphics[width=.48\textwidth]{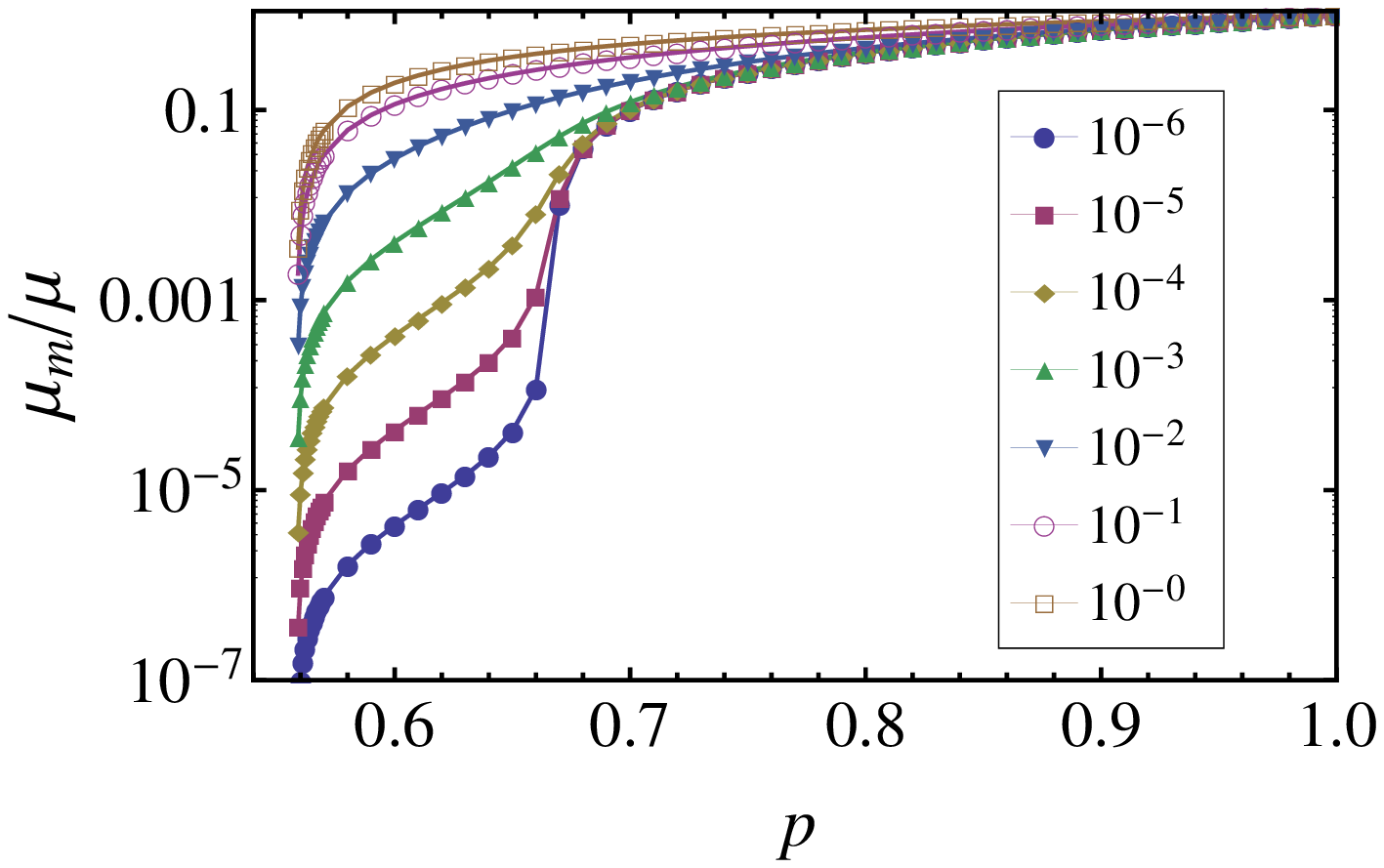}}
     \subfigure[]{\includegraphics[width=.45\textwidth]{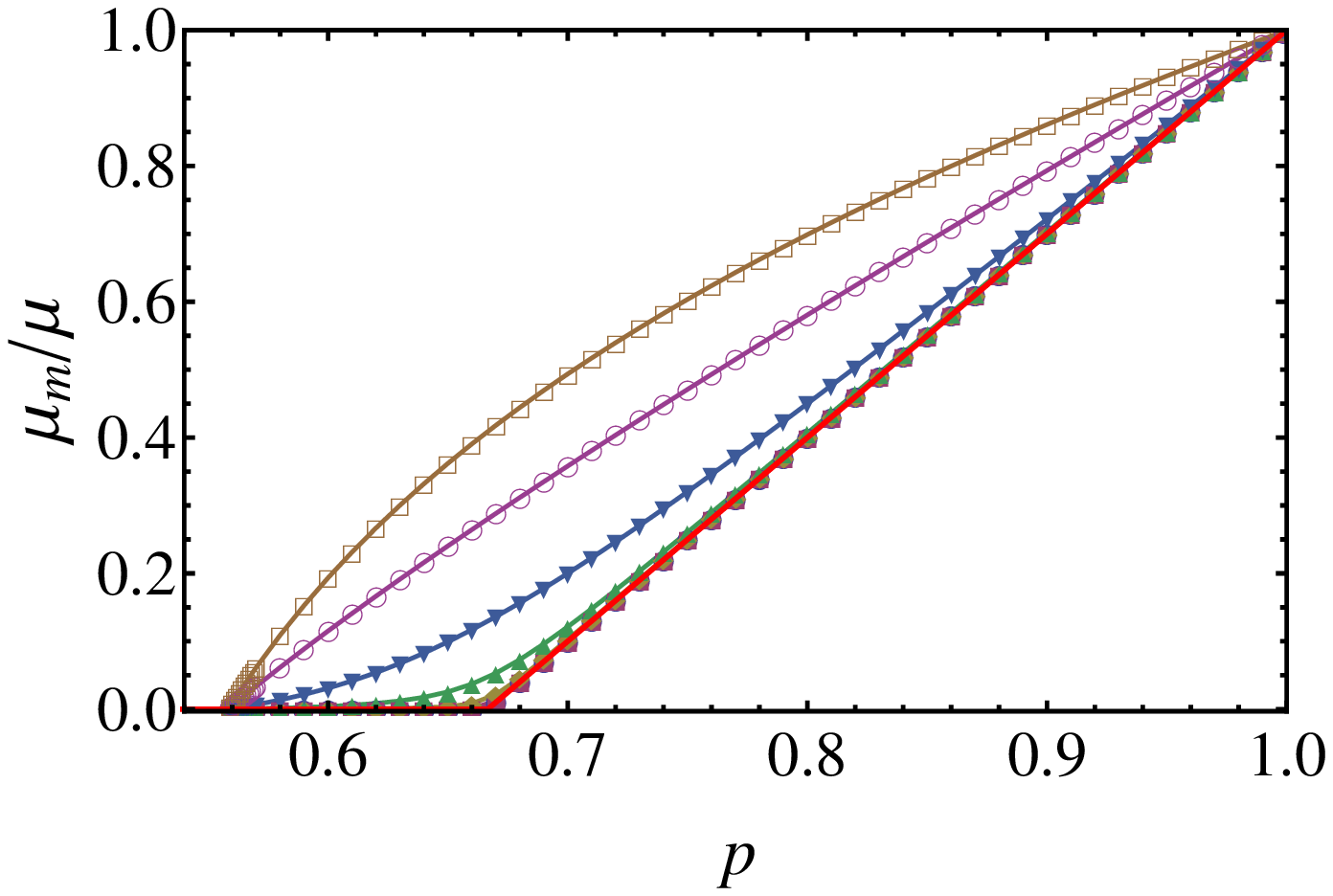}}
        \caption{(color online) (a) Semi-log plot for the EMT solution $\sm/\mu=G/G_0$ as a function of
        $\Prob$ for $\s=1$ and $\b=1,10^{-1},10^{-2},10^{-3},10^{-4},10^{-5},10^{-6}$
        from top to bottom, as indicated in the legend. (b) Linear plot of $\sm/\mu$ as a function
        of $\Prob$, with parameters and color code the same as in (a).  The red solid
        line indicates $\sm$ for the case of a central force triangular lattice ($\b=0$). 
        Here and in Figs.~\ref{FIG:CPASOLUBL} to \ref{FIG:SCAL}, we have set $\a=1$. The
        contents of (a) appeared in a different form in Ref.~\cite{BroederszMac2011}.}
\label{FIG:CPASOLUK}
\end{figure}

\begin{figure}[h]
    \centering
     \subfigure[]{\includegraphics[width=.45\textwidth]{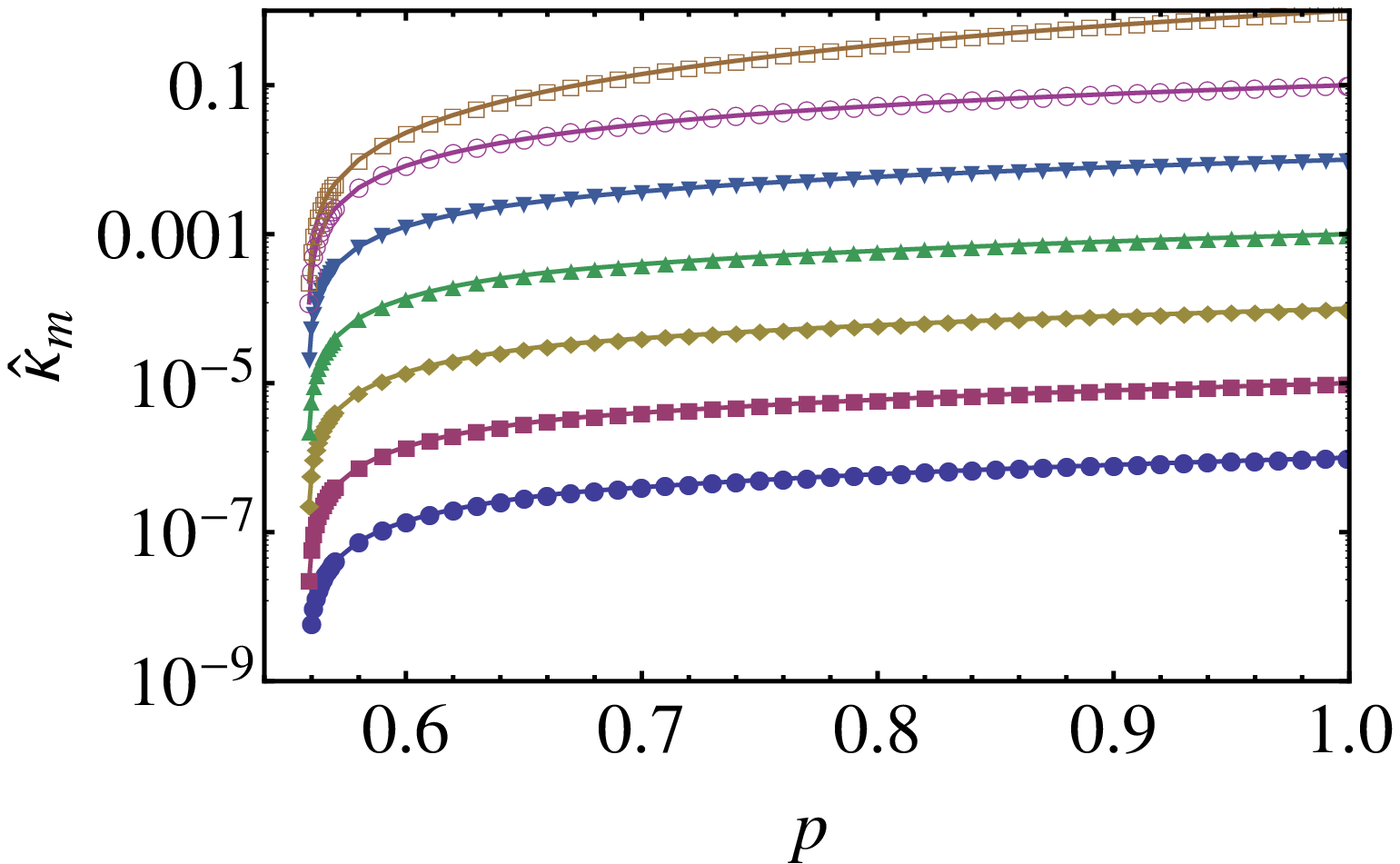}}
     \subfigure[]{\includegraphics[width=.45\textwidth]{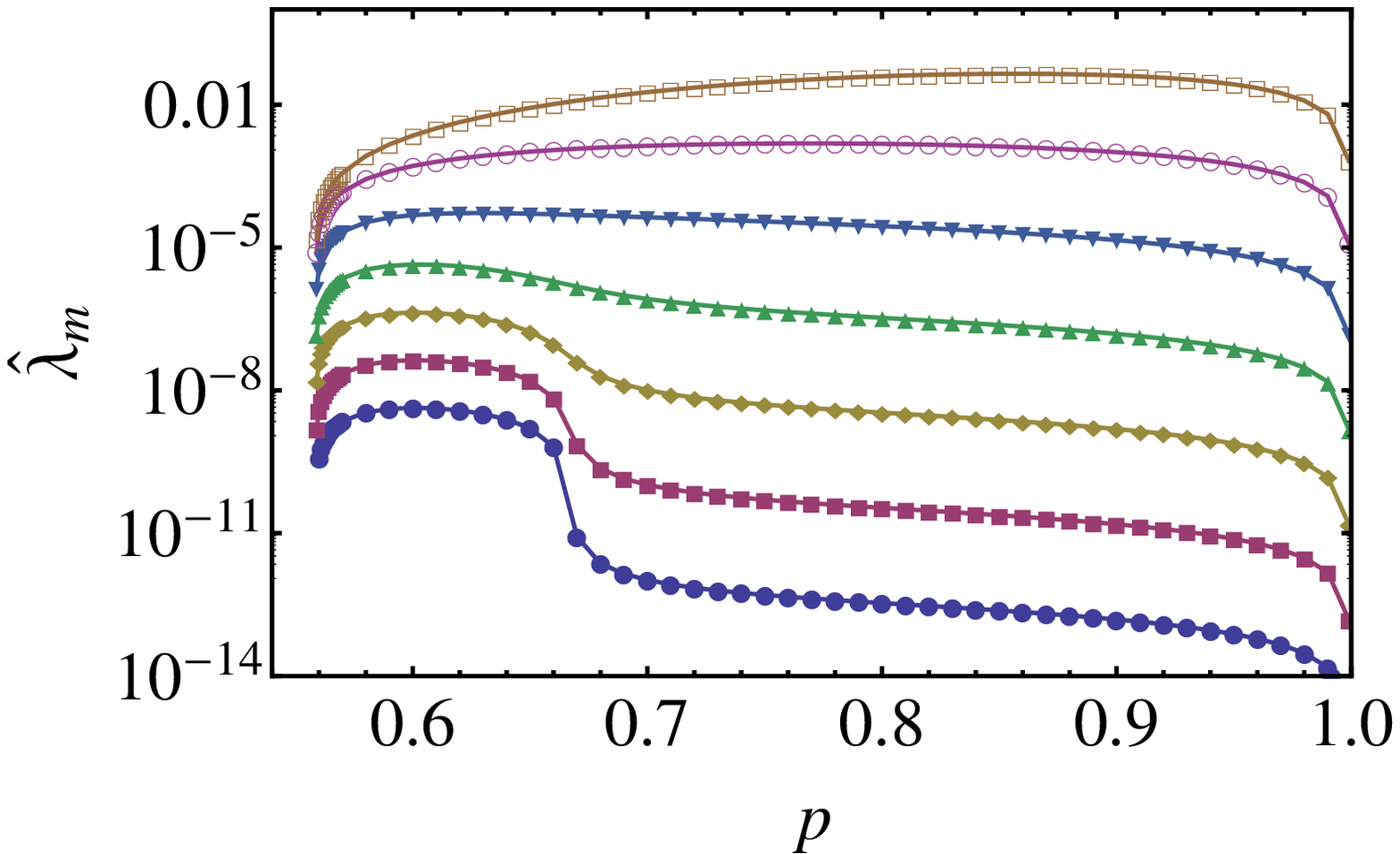}}
        \caption{(color online) The EMT solution for $\bm$ expressed in terms of the dimensionless combination $\hat{\kappa}_m=\bm/(\s a^2)$ (a) and $\lm$ expressed in terms of the dimensionless combination $\hat{\lambda}_m=\lm/(\s a^2)$(b).
        Parameters and color code are the same as in Fig.~\ref{FIG:CPASOLUK}.
        }
\label{FIG:CPASOLUBL}
\end{figure}

Numerical solutions to Eqs.~(\ref{EQ:AsymEqOne}) to
(\ref{EQ:AsymEqThe}) for any value of $\kappa/\mu$ are easily
calculated, and the results for the effective medium elastic
parameters are plotted in
Figs.~\ref{FIG:CPASOLUK},~\ref{FIG:CPASOLUBL}. There are
several properties of these plots that are worthy of note:
\begin{enumerate}
\item $\mu_m$ vanishes at the $CF$ Maxwell rigidity
    threshold $\PCF = 2/3$ when $\kappa=0$ and at $p=\Pb =
    0.56$ for {\em all} $\kappa
    >0$.  Simulations of the same model yield a
    slightly smaller value of  $\PCF\simeq 0.659$
    \cite{BroederszMac2011,ArbabiSah1995} and a
    considerably smaller $\Pb\simeq
    0.445$~\cite{BroederszMac2011}. [Using a variation
    of the Maxwell floppy mode count, we estimated the
    rigidity threshold in presence of filament bending
    stiffness and obtained $\Pb\simeq 0.448$ in good
    agreement with simulation results.  This calculation
    has been reported in the Supplementary Information of
    Ref.~\cite{BroederszMac2011}.]
\item$\mu_m$ increases with $\kappa$ for all $p
    >\Pb$.
\item For small $\kappa/(\mu \a^2)$, there is an
    interesting and nontrivial crossover near $\PCF$, which
    follows the analytic solution, Eq.~(\ref{EQ:gOne}), to
    the EMT equations, whereas for large $\kappa/(\mu
    \a^2)$, memory of the CF threshold is effectively lost
    and $\mu_m$ rapidly reaches value near its saturation
    value $\mu$ for $p>\Pb$.
\item $\kappa_m$ vanishes as $p \to \Pb$ and rises smoothly
    to its saturation value $\kappa$ without any evidence
    of crossover behavior near $\PCF$.
\item $\lambda_m$ vanishes at $\Pb$ and in the undiluted
    lattice ($p=1$), which it must by construction. It
    exhibits crossover behavior near $p=\PCF$ for small
    $\kappa/(\mu \a^2)$.
\end{enumerate}

\begin{figure}%[t]
    \centering
     \includegraphics[width=.45\textwidth]{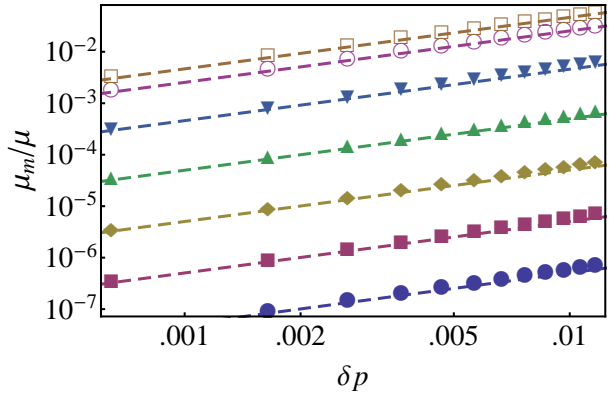}
        \caption{(Color online) Asymptotic solution (dashed lines) and numerical solutions
        (data points) of $\sm/\mu=G/G_0$ near $\Pb$.  Parameters and color code are the same as
        in Fig.~\ref{FIG:CPASOLUK}.
        }
\label{FIG:Pb}
\end{figure}

\begin{figure}%[h]
    \centering
     \subfigure[]{\includegraphics[width=.43\textwidth]{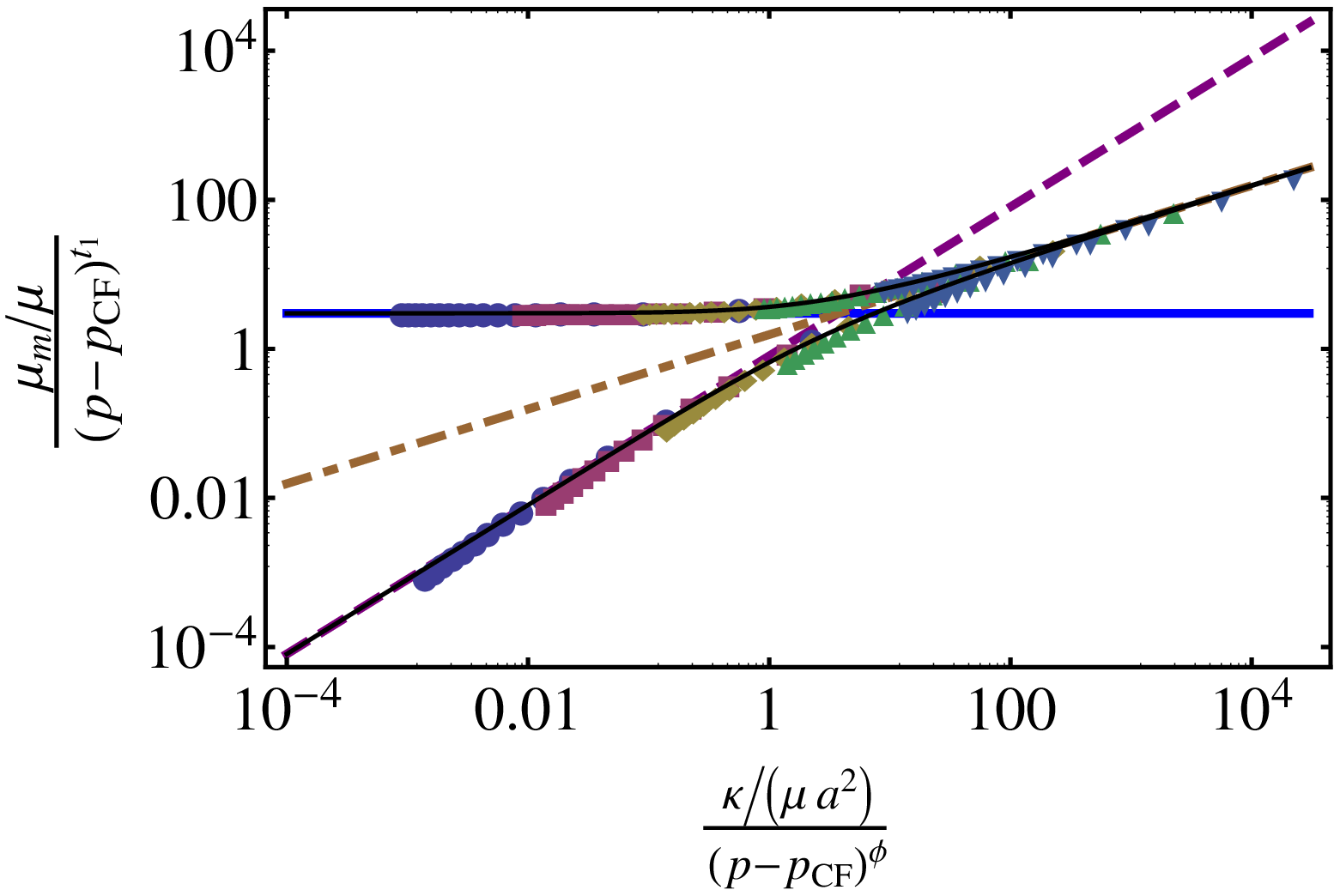}}
     \subfigure[]{\includegraphics[width=.43\textwidth]{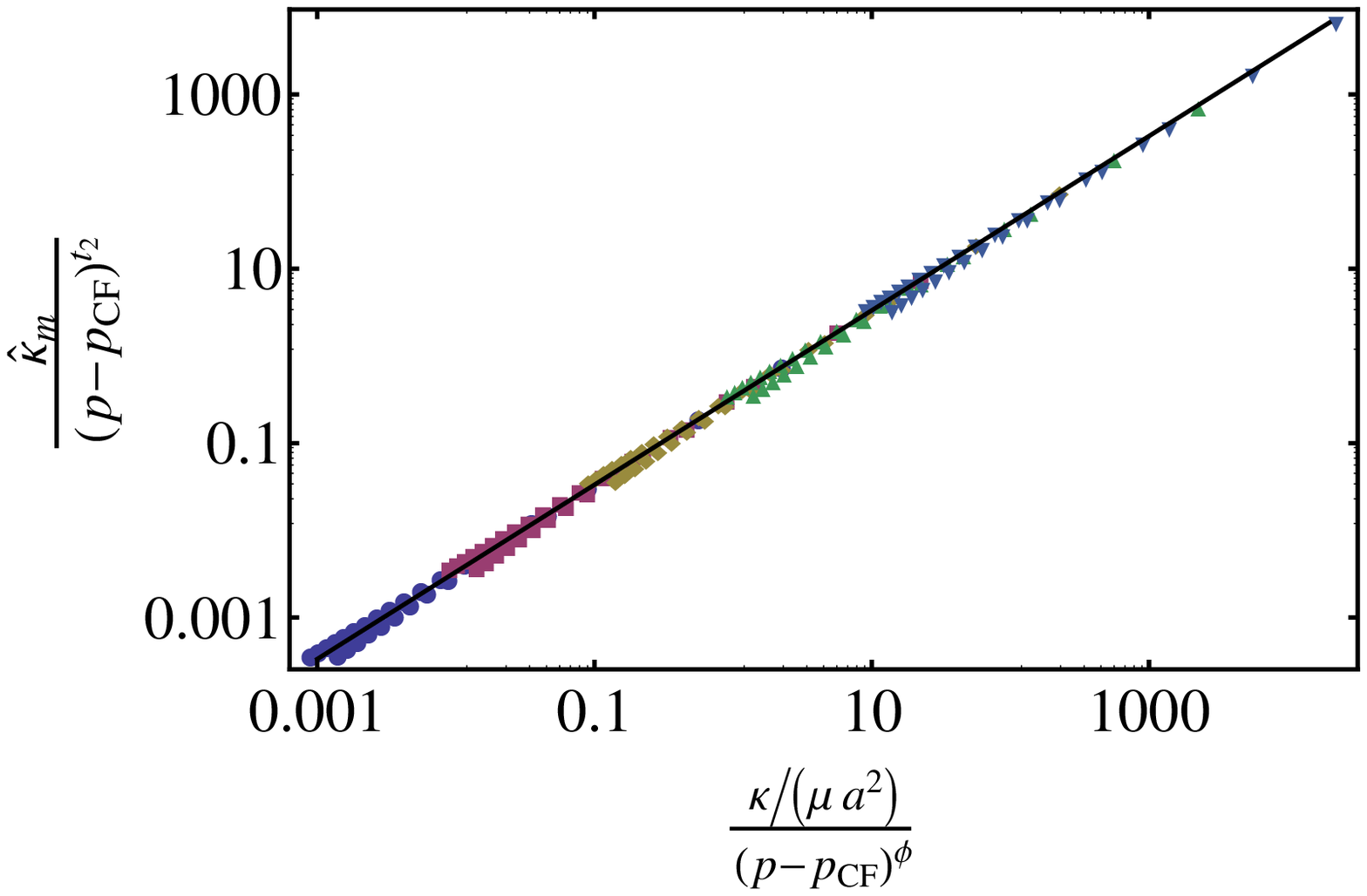}}
     \subfigure[]{\includegraphics[width=.43\textwidth]{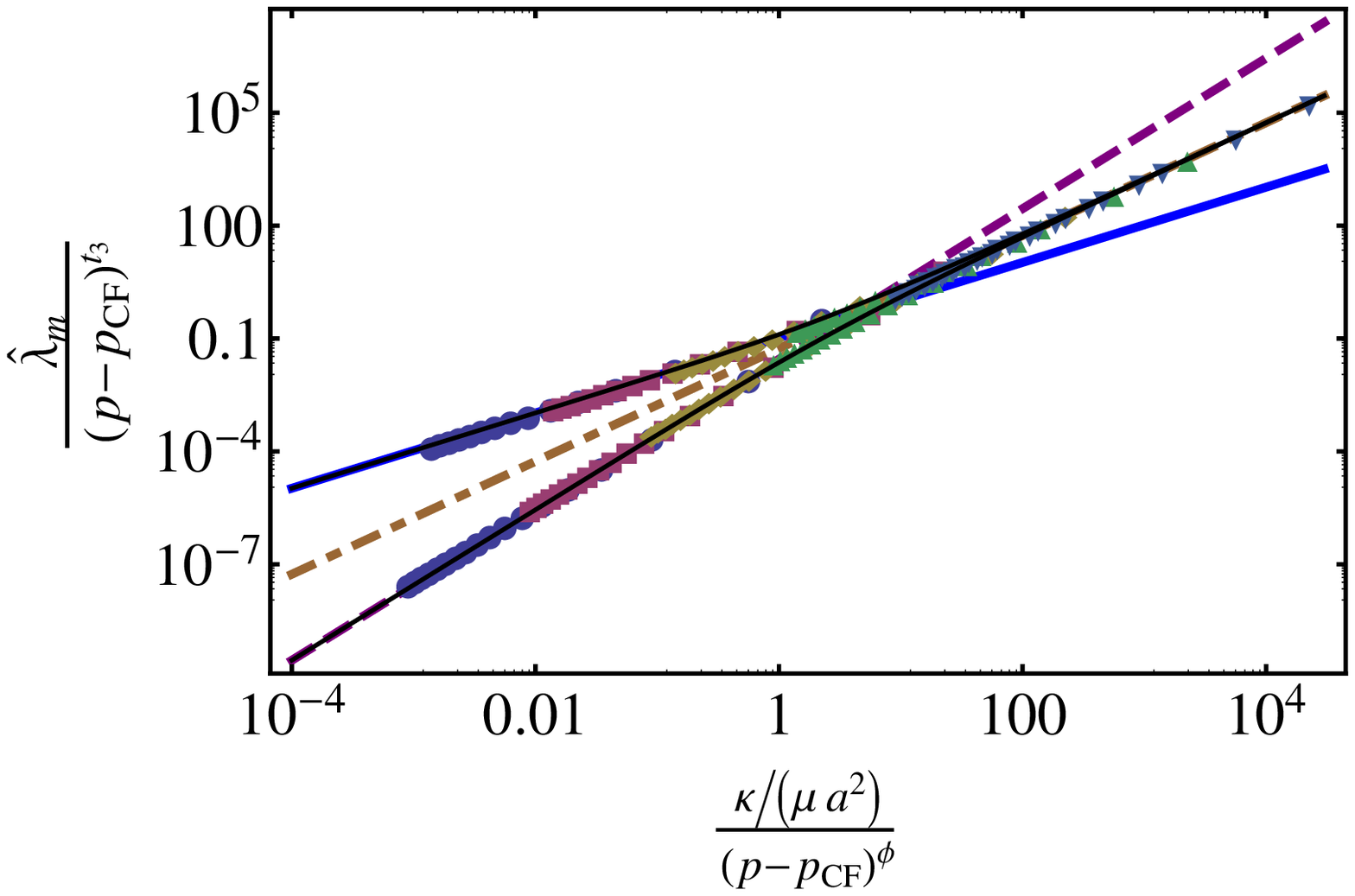}}
        \caption{(color online) (a), (b), and (c) show rescaled plots of the EMT solutions
        $\sm/\mu$, $\bm$ and $\lm$ using the scaling forms~(\ref{EQ:Scal})
        for $\s=1$ and $\b=10^{-2},10^{-3},10^{-4},10^{-5},10^{-6}$, with color
        code the same as in Fig.~\ref{FIG:CPASOLUK}, and exponents taking the value
        as in Eq.~(\ref{EQ:Expo}). The thin black lines represent the asymptotic forms
        of Eq.~(\ref{EQ:ScalFunc}) for small $\b$.  The brown dash-dotted lines,
        the thick blue solid lines, and the purple dashed lines plot the functional
        form of $\sm$ obtained in the crossover, the stretching-dominated, and the
        bending-dominated regimes of Eq.~(\ref{EQ:gOne}), respectively. The
        contents of (a) appeared in a different from in Ref.~(\cite{BroederszMac2011})
        }
\label{FIG:SCAL}
\end{figure}

In Figs.~\ref{FIG:Pb} and \ref{FIG:SCAL} we respectively plot
our numerical solutions to the EMT equations near $\Pb$ and
$\PCF$ using the analytic scaling forms of
Eqs.~(\ref{EQ:AsymEqOne}) and (\ref{EQ:gOne}). As required the
numerical solutions agree with the analytic ones.

\section{Discussion \label{SEC:DISCUSSION}}

Two other approaches, one by Das {\em et al.}
\cite{Das2007,Das2012} and one by Wyart {\em et al.}
\cite{WyartMah2008}, produce results similar to ours, and below
we briefly review compare them to ours. References
\cite{Heussinger2006} and \cite{HeussingerFre2007a}, which
develop an EMT for fiber networks like the Mikado model with a
maximum coordination number of four, do not consider rigidity
development on a triangular lattice with a maximum coordination
number of six, and we will not discuss them further.

Stretching forces are easily described by CF springs, which
reside on bonds, each of which can have a distinct spring
constant.  Bending forces, on the other hand, couple angles on
neighboring $NN$ bonds, or equivalently $NNN$ sites along a
filament to the site between them via \emph{phantom} $NNN$
bonds. Because removing one $NN$ bond from a pair defining a
phantom $NNN$ bending bond effectively removes that phantom
bond, bending and stretching are not independent in the diluted
lattice. This presents real challenges for the development of a
consistent bend-stretch EMT.

Our approach to this problem appeals to the underlying polymer
nature of our model in which constituent polymers are endowed
with local stretching and bending moduli $\mu$ and $\kappa$. We
can modify these moduli along any bond. Different stretch
moduli lead to independent effective CF stretch force constants
$k_{||}=\mu/\a$ for each bond. Modification of the bending
modulus $\kappa$ on a given $NN$ bond, however, modifies the
bend force constant $k_{\perp}=\kappa/\a^3$ for {\em both}
phantom $NNN$ bonds that that $NN$ bond partially defines in
the manner described above. With this approach, we develop a
consistent EMT that includes the statistical correlation
between bend and stretch.

Das {\em et al.} begin by ignoring the correlation between real
$NN$ bonds and phantom $NNN$ bonds and assume that a stretch
spring on a given $NN$ bond can be removed without affecting
the bending energy on the phantom $NNN$ bonds that include that
$NN$ bond and that bending springs on the phantom $NNN$ bonds
can be removed without affecting the stretch springs on the two
bonds that define the $NNN$ bond. In other words, the phantom
bond is effectively elevated to a real bond with existence
independent of the underlying $NN$ bonds. In general, the $NNN$
bonds can be present with with an arbitrary probability $q$ and
absent with probability $1-q$. To provide an approximate
description of the constraint that the phantom bond does not
exist unless both of the $NN$ bonds defining it are present,
Das {\em et al.} assign a probability $q=p^2$ ($p$ is the
probability that a $NN$ bond is occupied) to the occupancy of a
$NNN$ bending bond, but continue to treat the $NN$ and $NNN$
bonds as statistically independent. Again the result is a set
of closed self-consistent equations for $\mu_m$ and $\kappa_m$,
which we analyze in our formulation of EMT in
App.~\ref{APP:Comp}.

Both approaches yield $\PCF = 2/3$ in good agreement with
numerical estimates
\cite{WangAdl1992,Sahimi1998,BroederszMac2011}, which yield
$\PCF$ of order $0.64$ or $0.65$. Our approach yields a value
for $\Pb$ (0.56) that is well above that (0.445) observed in
simulations \cite{BroederszMac2011} whereas that of
Ref.~\cite{Das2012} yields a value ($\Pb = 0.457$) in good
agreement with simulations.  The latter method
produces results in better agreement with simulations over the
entire range of values of $p$ than does ours if no
approximations to the EMT equations are used in the numerical
evaluation of the shear modulus [See App.~\ref{APP:Comp}]. It
is not clear to us why this is so. Both approaches yield a
nontrivial bend-stretch crossover, with the same algebraic form
but with slightly different parameters [See
App.~\ref{APP:Comp}] in the vicinity of $\PCF$ in qualitative
agreement with simulations.

In Ref.~\cite{WyartMah2008}, Wyart \emph{et al.} consider
random off-lattice elastic networks derived from
two-dimensional packings of spheres \cite{OHern2003} with a
cordination number above the Maxwell CF isostatic limit of
$z=4$ in which CF springs are assigned to each sphere-sphere
contact.  They use numerical simulations to study the nonlinear
relation between shear stress $\sigma$ and shear strain
$\gamma$ as springs are cut, thereby reducing $z$, and they
find a scaling relation $\sigma = \gamma |\delta z|
f(\gamma/|\delta z|)$, where $\delta z = z - 4$, $f(x)\to {\rm
const.}$ for $x\to 0^+$, $f(x)\to 0 $ for $x\to 0^-$, and $f(x)
\sim x$ for $x\to \infty$.  This scaling form predicts $\sigma
\sim \gamma$ for $\delta z \gg \gamma >0$, $\sigma = 0$ for
$\delta z \ll -\gamma <0$, and $\sigma \sim \gamma^2$ for
$\gamma \gg \delta z$.  Reference \cite{WyartMah2008} then
provides a theoretical justification for this behavior based
upon the existence of a plateau in the density of states
\cite{Silbert2005} above $\omega^* \sim \delta z$ and
reasonable assumptions about statistical independence of
eigenvectors associated with different normal modes in the
isostatic network \cite{Maloney2006} and about the nature of
nonaffine response of nearly isostatic systems. Finally, they
extend this line of reasoning to nearly isostatic systems with
extra weak bonds and find three regimes of elastic response
that are identical to those we identify in Eqs.~(\ref{EQ:Scal})
to (\ref{EQ:ScalFunc}) if the weak bonds are of a bending type.

To summarize we developed an effective-medium theory that can
include bending energy of filaments, and we used it to study
the development of rigidity of a randomly diluted triangular
lattice with central force springs on occupied bonds and
bending forces between occupied bond pairs along a straight
line. We obtained a rigidity threshold for positive bending
stiffness and a crossover, controlled by the isostatic point of
the central force triangular lattice, characterizing
bending-dominated, stretching-dominated, and stretch-bend
coupled elastic regimes.

\noindent
{\it Acknowledgments\/}---%
We are grateful to  C. P. Broedersz and F.C. MacKintosh for
many stimulating and helpful discussions. This work was
supported in part by the National Science Foundations under
grants DMR-0804900 and DMR-1104707 and under the  Materials
Research Science and Engineering Center DMR11-20901 .

\appendix

\section{Discretization of a continuous rod}\label{APP:COMP}

In this appendix, we will derive the discretized bending energy
for an inhomogeneous rod from the continuous bending energy in
of Eq.~(\ref{EQ:Rod}).  We divide the rod into bonds of length
$\a$ whose endpoints are at nodes $i$ (which coincide with
vertices of our lattice) as shown in
Fig.~\ref{FIG:discrete-fil}. Segment $i$, which lies between
nodes $i-1$ and $i$, is endowed with a bending modulus
$\kappa_i$, and the angle at its center is constrained to be
$\theta_i$. Within each segment $i$, the angle $\theta_i(t)$
with $-\a/2 <t <\a/2$ ]i.e., within segment $i$, $t= s-i
a$] minimizes the bending energy in that segment and satisfies
the equation $d^2 \theta_i (t)/dt^2 = 0$ subject to the
boundary conditions (BCs) for each $i$
\begin{align}
(1)\,\,\,&\theta_i(t=0) &= &\theta_i ,  \nonumber\\
(2)\,\,\,& \theta_{i+1}(-\a/2) & = &\theta_{i}(\a/2) , \nonumber\\
(3)\,\,\, & \kappa_{i} \left. \frac{d \theta_{i}}{ds}\right|_{t=\a/2} &
= &\kappa_{i+1}\left. \frac{d \theta_{i+1}}{ds}\right|_{ts=-\a/2}  .
\label{EQ:A1}
\end{align}
BC (1) is the constraint that $\theta(t)$ take on the value
$\theta_i$ at the center of bond $i$; BC(2) is the condition
that $\theta(t)$ be continuous at node $i$; and BC(3) is the
condition that the torque on node $i$ be zero. Thus, within
segment $i$,
\begin{equation}
\theta_i(t) =
\begin{cases}
    \theta_i + A_i^- s & \mbox{if }  -\a/2 < t<0 \\
    \theta_i + A_i^+ s & \mbox{if }  \,\,0<t<\a/2 .
\end{cases}
\label{EQ:A2}
\end{equation}
This form immediately satisfies boundary condition $(1)$.
Boundary condition $(2)$ requires
\begin{equation}
\theta_{i} + \frac{\a}{2} A_{i}^+ = \theta_{i+1}
-\frac{\a}{2} A_{i+1}^- ,
\label{EQ:BC2}
\end{equation}
and boundary condition $(3)$ requries
\begin{equation}
\kappa_i A_i^+ = \kappa_{i+1} A_{i+1}^- .
\label{EQ:BC3}
\end{equation}
The solution of  Eq.~(\ref{EQ:BC2}) and (\ref{EQ:BC3}) for
$A_i^+$ and $A_{i+1}^-$ is
\begin{equation}
A_i^+=\frac{2}{\a} \frac{\kappa_{i+1}}{\kappa_i + \kappa_{i+1}}(\theta_{i+1}-\theta_i )
= \frac{\kappa_{i+1}}{\kappa_i} A_{i+1}^-.
\label{EQ:A_i}
\end{equation}
With this result, we can calculate the bending energy of the
segment running from the midpoint of bond $i$ to the midpoint
of bond $i+1$:
\begin{align}
E_i & = \frac{1}{2} \kappa_{i}\int_0^{\a/2}\left(\frac{d \theta_{i}}{ds}\right)^2
+ \frac{1}{2} \kappa_{i+1}\int_{-\a/2}^0\left(\frac{d \theta_{i+1}}{ds}\right)^2 \nonumber\\
& = \frac{1}{2}\frac{\a}{2}\left[\kappa_{i} (A_{i}^+)^2 + \kappa_{i+1} (A_{i+1}^-)^2\right] \nonumber\\
&= \frac{1}{2 \a} \kappa_i^{\text{eff}}(\Delta \theta_i)^2 ,
\end{align}
where $\Delta \theta_i = \theta_{i+1}-\theta_i$ and
\begin{equation}
\kbar_i= \frac{2 \kappa_{i+1} \kappa_i}{\kappa_{i+1} + \kappa_i} .
\end{equation}
When $\kappa_i= \kappa_{i+1}$, this reduces to $\kappa_i$.  The
total energy, apart from boundary terms, which we ignore, is
then $E= \sum_i E_i$.  When all $\kappa_i$ are equal, this is
indeed exactly the discretized  form that we use.  If the
bending modulus $\kappa_i$ on segment (bond) $i$ differs from
the modulus $\kappa$ on all of the other bonds, then the
bending energies associated with site $i-1$ and $i$ will have
an ``effective" modulus $2\kappa_i \kappa/(\kappa_i + \kappa )$
in agreement with our EMT treatment.

It is instructive to verify that the continuum and the
discretized theory give the same result for a particular
inhomogeneous $\kappa$. For simplicity, we consider a filament
of length $L$ whose left and right ends coincide with bond
centers (rather than nodes) at positions $t=0$ and $t=L$,
respectively. There are thus $N-1$ contiguous bonds of length
$\a$ terminated by two half bonds of length $\a/2$.  We assume
that the bending modulus is equal to $\kappa_1$ in regions I
defined by $0\leq t < t_p = [p-(1/2)]\a$ and to $\kappa_2$ in
regions II defined by $t_p< s<\leq L$, and we assign boundary
conditions that $\theta(0) = 0$ and $\theta(L) = \Theta$.
Consider first the continuum case.  $d^2 \theta(t)/dt^2 = 0$ in
both regions I and II, and as a result the solutions for
$\theta$ in these two regions that satisfy the boundary
conditions are, respectively, $\theta_1 = B_1(t/\a)$ and
$\theta_2 = \Theta + B_2 [(t-L)/\a]$.  The additional boundary
conditions are that $\theta$ and $\kappa d\theta/dt$ be
continuous at $t=t_p$, implying
\begin{eqnarray}
(B_1/\a) t_p & = & \Theta + (B_2/\a) (t_p - L)  , \nonumber \\
\kappa_1 B_1/\a & = & \kappa_2 B_2/\a .
\end{eqnarray}
These equations are easily solved for $B_1$ and $B_2$:
\begin{equation}
\b_1 = \frac{\kappa_2 \Theta}{(\kappa_2 - \kappa_1) (t/\a) + \kappa_1(L/\a)}
= \frac{\kappa_2}{\kappa_1} B_2 .
\end{equation}

In the discrete case, nodes $i=1$ to $p-1$ have bending energy
$(1/2)\kappa_1(\theta_{i+1}-\theta_i)^2$, nodes $i=p+1$ to
$(N-1)$ have bending energy
$(1/2)\kappa_2(\theta_{i+1}-\theta_i)^2$, and site $p$ has
bending energy
$(1/2)\overline{\kappa}(\theta_{p+1}-\theta_p)^2$.  The
equations for $\theta_i$, $i=1, \cdots N-1$ are thus,
\begin{align}
\frac{dE}{d\theta_i} & = \kappa_b (2 \theta_i - \theta_{i+1} - \theta_{i-1} )=0;
\,\,\,\, 1<i<p-1 \nonumber\\
\frac{dE}{d \theta_{p-1}}& =\kappa_b (\theta_{p-1}-\theta_{p-2}) +
\overline{\kappa} (\theta_{p-1} - \theta_p ) =0 ; \nonumber\\
\frac{dE}{d \theta_p} & = \overline{\kappa} (\theta_p - \theta_{p-1})
+\kappa_a ( \theta_p - \theta_{p+1})=0 ;\nonumber\\
\frac{dE}{d \theta_i} & = \kappa_a (2 \theta_i - \theta_{i+1} - \theta_{i-1}) =0;
\,\,\,\,  p<i\leq N-2 .
\end{align}
These linear difference equations subject to the boundary
conditions, $\theta_0 = 0$ and $\theta_{N} = \Theta$ are solved
by setting $\theta_i = D_1 i$ in region I ($0\leq i \leq p$)
and $\theta_i = \Theta + D_2 (i-N)$ in region II ($p< i \leq
N$). The equilibrium equations for $\theta_p$ and
$\theta_{p-1}$ are
\begin{align}
[\kappa_1+ \overline{\kappa}(p-1)] D_1 - \overline{\kappa} (p-n) D_2
& = & \,\,\overline{\kappa}\, \Theta  , \nonumber \\
-\overline{\kappa}(p-1) D_1 + [\overline{\kappa}(p-N)-\kappa_2 ]D_2
& = & - \overline{\kappa} \,\Theta .
\end{align}
These equations, along with the relation $\overline{\kappa} = 2
\kappa_1 \kappa_2/(\kappa_1+\kappa_2)$, yield $D_1=B_1$ and
$D_2 = B_2$  verifying that the discrete and continuum
solutions agree.

Finally, we need to specify the relation between angles
$\theta_i$ and the vertical displacements $h_i$ (i.e.,
$u_i^{\perp}$).  Let $h'_i$ be the height at the center of bond
$i$. To linear order in continuum theory, $dh(t)/ds =
\theta(t)$. Integration of this equation [using
Eq.~(\ref{EQ:A2})] then yields
\begin{eqnarray}
h_i - h'_i & =& \int_{0}^{\a/2} \theta_i(t) dt =(\a/2) \theta_i+\frac{a^2}{8} A_i^+\nonumber\\
h_i'-h_{i-1} & = & \int_{-\a/2}^0 \theta_i(t) dt = (\a/2) \theta_i +\frac{a^2}{8} A_i^-
\end{eqnarray}
with the same convention as that of Eqs.~(\ref{EQ:A1}) and
(\ref{EQ:A2}). From Eq.(\ref{EQ:A_i}), $A_i^+ \propto
(\theta_{i+1}-\theta_i)/a$ and $A_i^- \propto (\theta_i -
\theta_{i-1})/a$.  Thus for slowly varying $\theta$ and small
$\a$, $\a^2 A_i^+$ and $\a^2 A_i^-$ can be ignored relative to
$\a \theta_i$.  This is true whether or not $\kappa$ changes
from bond to bond. The result is that the slope of $h(s)$
within bond $i$ is simply $\theta_i$, and
\begin{equation}
\frac{h_i - h_{i-1}}{a} \approx \theta_i + \frac{1}{8} a (A_i^+ - A_i^- ) .
\end{equation}

%%%%%%%%%%%%%%%%%%%%%%%%%%%%%%%%%%%%%%%%%%
\section{The dynamical matrix and the phonon
Green's function of the effective medium}\label{APP:DM} From
Eq.~(\ref{EQ:DMCons}), it is straightforward to calculate the
components of the dynamical matrix of the effective medium:
\begin{align}
	\DM = \left(
		\begin{array}{cc}
		D_{xx} & D_{xy} \\
		D_{yx} & D_{yy}
		\end{array}
	\right) ,
\end{align}
where
\begin{widetext}
\begin{align}
	D_{xx} =& \sm \left\lbrack
		3-2\cos q_x -\cos (q_x/2)\cos(\sqrt{3}q_y/2)
	\right\rbrack
	+ 3\bm \left\lbrack
		3-4\cos (q_x/2)\cos(\sqrt{3}q_y/2) + \cos (q_x)\cos(\sqrt{3}q_y)
	\right\rbrack \nonumber\\
	&+ 3\lm \left\lbrack
		-4+7\cos (q_x/2)\cos(\sqrt{3}q_y/2) -4 \cos (q_x)\cos(\sqrt{3}q_y)
		+\cos (3q_x/2)\cos(3\sqrt{3}q_y/2)
	\right\rbrack \nonumber\\
	D_{xy} =& \sqrt{3}(\sm-4\bm+7\lm)\sin(q_x/2)\sin(\sqrt{3}q_y/2)
	+\sqrt{3}(\sm-4\lm)\sin q_x \sin(\sqrt{3}q_y) \nonumber\\
	&+\sqrt{3}\lm \sin (3q_x/2) \sin(3\sqrt{3}q_y/2) \nonumber\\
	D_{yx}=&D_{xy} \nonumber\\
	D_{yy}=& 3\sm \left\lbrack
		1 -\cos (q_x/2)\cos(\sqrt{3}q_y/2)
	\right\rbrack \nonumber\\
	&+ \bm \left\lbrack
		9- 8\cos q_x +2\cos(2q_x)
		-4\cos (q_x/2)\cos(\sqrt{3}q_y/2) + \cos (q_x)\cos(\sqrt{3}q_y)
	\right\rbrack \nonumber\\
	&+ \lm \Big\lbrack
		-12+ 14\cos q_x -8\cos(2q_x)+2\cos (3q_x)
		+7\cos (q_x/2)\cos(\sqrt{3}q_y/2) \nonumber\\
		&\quad\quad -4 \cos (q_x)\cos(\sqrt{3}q_y)
		+\cos (3q_x/2)\cos(3\sqrt{3}q_y/2)
	\Big\rbrack .
\end{align}
\end{widetext}
The symmetry properties of the above components of $\DM$ will
determine which components of $\tGM$ are nonzero.
$D_{11}(q_x,q_y)$ and $D_{22}(q_x,q_y)$ are even under $q_x
\rightarrow -q_x$ and under $q_y\rightarrow -q_y$ whereas
$D_{12} (q_x,q_y)$ is odd under the same operations.

The effective medium phonon Green's function is the negative of
the inverse of $\DM$
\begin{align}
	\GF^{(m)} =& -(\DM)^{-1}\nonumber\\
	=& (\textrm{Det}\DM)^{-1} \left(
		\begin{array}{cc}
		D_{yy} & -D_{xy} \\
		-D_{yx} & D_{yy}
		\end{array}
	\right) .
\end{align}
The determinant $\textrm{Det}\DM= D_{xx}D_{yy} - D_{xy}D_{yx}$
is even under $q_x \rightarrow -q_x$ and under $q_y\rightarrow
-q_y$, and thus $xx$ and $yy$ components of $\GF^{(m)}$ are
both even and the $xy$ component of $\GF^{(m)}$ is odd under
$q_x \rightarrow -q_x$ and under $q_y\rightarrow -q_y$. With
this information and the properties of $\vB^{s}_{n,\vq}$ and
$\vB^{b}_{n,\vq}$, we can infer which components of $\tGM$ are
zero and which are equal to each other.  First consider the
$12$ and $13$ components, which from Eq.~(\ref{EQ:DEFtG}) are
given by
\begin{eqnarray}\label{EQ:tGM123}
(\tGM)^{12} & = & \frac{1}{N}\sum_{\vq} \vB^{s}_{1,-\vq} \GM_{\vq} \vB^b_{1,\vq}\\
(\tGM)^{13} & = & \frac{1}{N}\sum_{\vq} \vB^{s}_{1,-\vq} \GM_{\vq} \vB^b_{4,\vq} .
\end{eqnarray}
$\vB^{s}_{1,-\vq}$ is a vector parallel to the $x$-axis (i.e.,
to  $\uvepara_1$), whereas both $\vB^b_{1,\vq}$ and
$\vB^b_{4,\vq}$ are parallel to the $y$-axis (i.e. to
$\uveperp_1$).  In addition, $\vB^{s}_{1,-\vq}$,
$\vB^b_{1,\vq}$, and $\vB^b_{4,\vq}$ are all even under $q_y
\rightarrow - q_y$.  Thus the integrands in
Eq.~(\ref{EQ:DEFtG}) are equal to $G_{xy}$ times a function
even under $q_y \rightarrow - q_y$.  Since $G_{xy}$ is odd
under this operation, both integrals vanish, and the $12$ and
$13$ components of $\tGM$ vanish by symmetry. There are no
symmetry operations that make the other components of $\tGM$
vanish, but the relation $\vB^{b}_{1,-\vq}
\vB^b_{1,\vq}=\vB^b_{4,-\vq}\vB^b_{4,\vq}$ sets the the $22$
and $33$ component of $\tGM$ equal to each other and leads to
Eq.~(\ref{EQ:tGF}) for $\tGM$.

\section{Comparison with EMT results obtained using methods in Ref.~\cite{Das2007,Das2012}}
\label{APP:Comp}
In this section we derive the Das' EMT equations from our
approach (by changing some assumption as detailed below) and
calculate it for the triangular lattice.  We also compare it to
both our EMT and simulation results.

We start from the same effective medium dynamical matrix $\DM$
as we defined in the main text ($\DM$ with $\sm, \bm$ but
$\lm=0$), but we make different assumptions about the changed
bond in the EMT.  In particular, the perturbative potential is
now
\begin{eqnarray}
	E_{V,\textrm{Das}}
	&=& \frac{1}{2}\frac{\ss - \sm}{\a}
	 (\vu_{\ell_1\ell_2}\cdot \uvr_{\ell_1\ell_2})^2 \nonumber\\
	&&+  \frac{1}{2} \frac{\bs-\bm}{\a^3}
    		 \theta_{\ell_1\ell_2\ell_3} ^2 ,
\end{eqnarray}
and the differences comparing with our version are: (i) there
is only 1 bending energy term $\theta_{\ell_1\ell_2\ell_3} ^2$
and the other term $\theta_{\ell_0\ell_1\ell_2} ^2$ is not
included, (ii) the bending stiffness is directly $\bs$ instead
of our composite one $\kappa_c(\bs)$, and (iii) there is no
$\lambda_m$ term.

The matrix form of $V$ in the space of $\{\vB_1^s , \vB_1^b\}$
is then (now $\vB_4^b$ is not relevant)
\begin{align}
	\tilde{\mathbf{V}}_{\textrm{Das}} = \left(
		\begin{array}{cc}
		(\ss-\sm)/\a & 0 \\
		0 & (\bs-\bm)/\a^3
		\end{array}
	\right)
\end{align}
Thus it is clear that
\begin{align}
	\tilde{\mathbf{T}}_\textrm{Das}
	= (\tilde{\mathbf{V}}_\textrm{Das}^{-1} - \tilde{\GF}^{(m)})^{-1}
\end{align}
is also a diagonal matrix (we have already proved in the text
that $\tilde{\GF}^{(m)}$ is diagonal due to symmetry).

Correspondingly the probability distribution is now
\begin{align}
	P_\textrm{Das}(\ss,\bs)
	=& \Big\lbrack p \delta(\ss-\s)+(1-p)\delta(\ss)\Big\rbrack \nonumber\\
	 &\times \Big\lbrack p^2 \delta(\bs-\b)+(1-p^2)\delta(\bs)\Big\rbrack  ,
\end{align}
with the distribution of $\ss$ and $\bs$ factorized.

Therefore the EMT matrix equation
\begin{align}
	\int d\ss d\bs P_\textrm{Das}(\ss,\bs) \tilde{\mathbf{T}}_\textrm{Das} (\ss,\bs) =0
\end{align}
decouples to two equations of the two diagonal elements (they
still share the same variables) that
\begin{align}\label{EQ:SCDAS}
	\frac{\sm}{\s} &= \frac{p-a^*}{1-a^*} \nonumber\\
	\frac{\bm}{\b} &= \frac{p^2-b^*}{1-b^*}
\end{align}
where
\begin{align}
	a^* &= \frac{\sm}{\a} \frac{1}{N}\sum_{\vq} \vB_{1,-\vq}^{s} \cdot \DM_{\vq}
	\cdot \vB_{1,\vq}^{s} \nonumber\\
	b^* &= \frac{\bm}{\a} \frac{1}{N}\sum_{\vq} \vB_{1,-\vq}^{b} \cdot \DM_{\vq}
	\cdot \vB_{1,\vq}^{b} .
\end{align}
which is exactly the equations from Mo's paper.

In contrast the distribution in our EMT is
\begin{align}
	P(\ss,\bs) = p \delta(\ss-\s) \delta(\bs-\b) + (1-p) \delta(\ss)\delta(\bs)
\end{align}
with the distribution of $\ss$ and $\bs$ correlated.  This is
more reasonable because they describe the same replaced bond.
Furthermore, $\bs$ affects two bending terms.

To summarize, the stretching bonds and \lq\lq bending
bonds\rq\rq  are treated as independent in the Das'  EMT,
whereas we model them as describing filament properties and
thus correlated.

From the definition of $\DM$ it is clear that
\begin{align}
	a^* +b^*
	&= \frac{2}{z}\, \textrm{tr} \Big\lbrace \frac{1}{N}\sum_{\vq}
	\big\lbrack \DM_{\vq} \big\rbrack^{-1} \DM_{\vq} \Big\rbrace \nonumber\\
	&= \frac{2d}{z} = 2/3
\end{align}
The self-consistency equation~\eqref{EQ:SCDAS} can be solved
numerically for any given $p$, $\s$, and $\b$.  In particular,
the rigidity threshold $\Pb$ can be solved analytically from
\begin{align}\label{EQ:SCDAS1}
	0 &= p-a^* \nonumber\\
	0 &= p^2-b^*
\end{align}
which leads to
\begin{align}
	\Pb = \frac{1}{2}\left( -1+\sqrt{1+\frac{8d}{z}}\right)
\end{align}
and for the triangular lattice it gives
\begin{align}
	\Pb \simeq 0.4574 .
\end{align}

\begin{figure}
	\includegraphics[width=.45\textwidth]{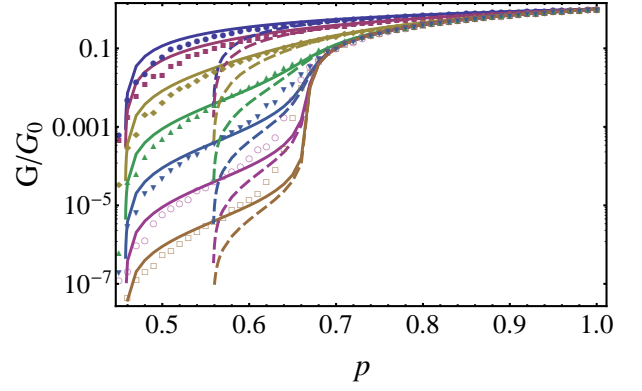}
	\caption{Comparison between the shear modulus (normalized by the shear
    modulus at $p=1$, so it is equivalent to $\mu_m/\mu$) obtained from numerical
    simulations from Ref.~\cite{BroederszMac2011}(data points), Das' EMT (solid lines), and our EMT (dashed lines).  Different colors mark different values of $\b$ with the same color code as in Fig.~\ref{FIG:CPASOLUK}, and from top to bottom the corresponding values of $\b$ are $1$, $10^{-1}, 10^{-2}, \cdots, 10^{-6}$.}
	\label{FIG:Compare}
\end{figure}

The EMT self-consistency equation~\eqref{EQ:SCDAS} can be
solved numerically, and we plot the results along with ours and
the simulation data from Ref.~\cite{BroederszMac2011} in
Fig.~\ref{FIG:Compare}. The curves calculated from
Eq.~\eqref{EQ:SCDAS} differ in detail from those presented in
Ref.~\cite{Das2012} because the latter reference used
approximate forms for $a^*$ and $b^*$ in its numerical
evaluations \cite{DasJen2012}

Near $\PCF$ we can also expand the Das' EMT solution to get the
asymptotic behaviors.  The functions $a^*$ and $b^*$ are
related to the integrals we defined via
\begin{align}
	a^* &= H_1(\bmr,0) \nonumber\\
	b^* &= \bmr H_2 \left(\bmr,0\right) .
\end{align}
At $\b=0$, because
\begin{align}
	H_1 (0,0) = \PCF, \quad H_2 (0,0)=0
\end{align}
it is straightforward to see that the Das' EMT lead to the same
central force solution as our EMT
\begin{align}
	\sm &= \s (\Prob-\PCF) \tilde{\Theta} (\Prob-\PCF) , \nonumber\\
	\bm &= 0 ,
\end{align}
where $\tilde{\Theta}$ is the Heaviside step function. For
small $\kappa>0$ we expand around small $\bmr=\bm/(\sm a^2)$.
We have already discussed the expansion of $H_1$ and $H_2$ at
this limit, and thus
\begin{align}
	a^* &=  \PCF + H_{1,1} \bmr , \nonumber\\
	b^* &= H_{1,1} \bmr .
\end{align}
Because $H_{1,1}$ is of order unity, $b^* \sim \bmr$ is very
small.  Therefore we can ignore the $b^*$ terms in the equation
of $\bm$ and get
\begin{align}
	\bm \simeq \b \Prob^2 ,
\end{align}
which differs from our EMT solution
\begin{align}
	\bm \simeq \b (2\Prob-1)
\end{align}
by the dependence on $\Prob$, but this difference is small and
does not show singularity near $\PCF$.  We can then plug this
solution back into the equation for $\sm$, which turns into a
quadratic equation similar to the equation for $\sm$ in our
EMT, with the only difference being the different $\Prob$
dependence of $\bm$.  We thus arrive at
\begin{align}
	\sm = \s \vert \Delta \Prob\vert \frac{3}{2} \left(
	\pm 1 +\sqrt{1-\frac{16\mathcal{A}}{27} \frac{\b}{a^2 \s \vert \Delta \Prob \vert^2}}
	\right)
\end{align}
which takes a very similar form to Eq.~\eqref{EQ:gOne}  just
with a different constant factor before $\b/(a^2  \s \vert
\Delta \Prob \vert^2)$.  Therefore the two EMTs produce the
same scaling behavior near $\PCF$.

%\bibliography{kagome_CPA_Refs-1}

%

\end{document}